\documentclass[reprint,
superscriptaddress,
nobibnotes,
amsmath,amssymb,
aps,
nolongbibliography,
]{revtex4-2}

\usepackage{dcolumn}
\usepackage{amssymb, esint}
\usepackage{physics}
\usepackage{bm, bbm}
\usepackage{graphicx}
\usepackage[colorlinks,allcolors=blue]{hyperref}
\usepackage{float}
\usepackage{xcolor}
\usepackage{url}
\usepackage[normalem]{ulem}

\def \br{\mathbf{r}}
\def \bA{\mathbf{A}}
\def \pol{\boldsymbol{\mathbf{{\varepsilon}}}}
\def \bk{\mathbf{k}}
\def \hbJp{\hat{\mathbf{J}}_{\rm{p}}}
\def \bJp{\mathbf{J}_{\rm{p}}}
\def \hbjp{\hat{\mathbf{j}}_{\rm{p}}}

\def \dpola{\boldsymbol{\tilde{\varepsilon}}_{\alpha}}
\def \dac{\hat{\tilde{a}}^{\dagger}}
\def \daa{\hat{\tilde{a}}}

\usepackage[nolist]{acronym}

\def\dunderline#1{\underline{\underline{#1}}}

\begin{document}

\title{Electron-Photon Exchange-Correlation Approximation for QEDFT}

\author{I-Te Lu}
\email{i-te.lu@mpsd.mpg.de}
\affiliation{Max Planck Institute for the Structure and Dynamics of Matter and Center for Free-Electron Laser Science, Luruper Chaussee 149, Hamburg 22761, Germany}

\author{Michael Ruggenthaler}
\email{michael.ruggenthaler@mpsd.mpg.de}
\affiliation{Max Planck Institute for the Structure and Dynamics of Matter and Center for Free-Electron Laser Science, Luruper Chaussee 149, Hamburg 22761, Germany}

\author{Nicolas Tancogne-Dejean}
\affiliation{Max Planck Institute for the Structure and Dynamics of Matter and Center for Free-Electron Laser Science, Luruper Chaussee 149, Hamburg 22761, Germany}

\author{Simone Latini}
\affiliation{Max Planck Institute for the Structure and Dynamics of Matter and Center for Free-Electron Laser Science, Luruper Chaussee 149, Hamburg 22761, Germany}
\affiliation{Department of Physics, Technical University of Denmark, 2800 Kgs. Lyngby, Denmark}

\author{Markus Penz}
\affiliation{Max Planck Institute for the Structure and Dynamics of Matter and Center for Free-Electron Laser Science, Luruper Chaussee 149, Hamburg 22761, Germany}
\affiliation{Department of Computer Science, Oslo Metropolitan University, 0130 Oslo, Norway}

\author{Angel Rubio}
\email{angel.rubio@mpsd.mpg.de}
\affiliation{Max Planck Institute for the Structure and Dynamics of Matter and Center for Free-Electron Laser Science, Luruper Chaussee 149, Hamburg 22761, Germany}
\affiliation{Center for Computational Quantum Physics (CCQ), The Flatiron Institute, 162 Fifth avenue, New York, NewYork 10010, United States of America}

\begin{abstract}


Quantum-electrodynamical density-functional theory (QEDFT) provides a promising avenue for exploring complex light-matter interactions in optical cavities for real materials. 
Similar to conventional density-functional theory, the Kohn-Sham formulation of QEDFT needs approximations for the generally unknown exchange-correlation functional.
In addition to the usual electron-electron exchange-correlation potential, an approximation for the electron-photon exchange-correlation potential is needed. 
A recent electron-photon exchange functional [\href{https://www.pnas.org/doi/abs/10.1073/pnas.2110464118}{C. Schäfer et al., Proc. Natl. Acad. Sci. USA {\bf{118}}, e2110464118 (2021)}], derived from the equation of motion of the non-relativistic Pauli-Fierz Hamiltonian, shows robust performance in one-dimensional systems across weak- and strong-coupling regimes. 
Yet, its performance in reproducing electron densities in higher dimensions remains unexplored.
Here we consider this QEDFT functional approximation from one to three-dimensional finite systems and across weak to strong light-matter couplings. 
The electron-photon exchange approximation provides excellent results in the ultra-strong-coupling regime. 
However, to ensure accuracy also in the weak-coupling regime across higher dimensions, we introduce a computationally efficient renormalization factor for the electron-photon exchange functional, which accounts for part of the electron-photon correlation contribution.
%
%
%
These findings extend the applicability of photon-exchange-based functionals to realistic cavity-matter systems, fostering the field of cavity QED (quantum electrodynamics) materials engineering.

\end{abstract}

\maketitle

\begin{acronym}
\acro{QEDFT}[QEDFT]{quantum-electrodynamical density-functional theory}
\acro{DFT}[DFT]{density-functional theory}
\acro{KS}[KS]{Kohn-Sham}
\acro{OEP}[OEP]{optimized-effective potential}
\acro{PF}[PF]{Pauli-Fierz}
\acro{px}[px]{electron-photon exchange}
\acro{pxLDA}[pxLDA]{electron-photon-exchange local-density approximation}
\acro{pc}[pc]{electron-photon correlation}
\acro{HO}[HO]{harmonic-oscillator}
\acro{EOM}[EOM]{equation of motion}
\acro{Mxc}[Mxc]{mean-field exchange-correlation}
\acro{JC}[JC]{Jaynes-Cummings}
\acro{HEG}[HEG]{homogeneous electron gas}
\acro{LDA}[LDA]{local-density approximation}
\acro{KLI}[KLI]{Krieger-Li-Iafrate}
\acro{pxc}[pxc]{electron-photon exchange-correlation}
\end{acronym}

\section{Introduction}
Optical cavities can, under specific conditions, enhance light-matter interaction without strong lasers or external pumping~\cite{ruggenthaler.flick.ea_2014,ruggenthaler.tancogne-dejean.ea_2018,hubener.degiovannini.ea_2021,flick.ruggenthaler.ea_2017,garcia-vidal.ciuti.ea_2021}, enabling precise control over materials properties~\cite{ebbesen_2016a,sentef.ruggenthaler.ea_2018a,ashida.imamoglu.ea_2020,latini.shin.ea_2021a,schlawin.kennes.ea_2022,jarc.mathengattil.ea_2023,kennes.rubio_2023}. 
Recent experimental advancements have allowed researchers to explore the strong, ultra-strong, or even deep-strong light-matter coupling regime~\cite{forn-diaz.lamata.ea_2019,friskkockum.miranowicz.ea_2019}.
On the other hand, in the realm of theoretical techniques for light-matter interactions~\cite{ruggenthaler.flick.ea_2014,ruggenthaler.tancogne-dejean.ea_2018,flick.rivera.ea_2018,riso.grazioli.ea_2023,ruggenthaler.sidler.ea_2023}, \ac{QEDFT} stands out as an efficient and accurate approach for realistic materials~\cite{ruggenthaler.flick.ea_2014,tokatly_2013a}. 
Exact \ac{QEDFT} treats electrons and photons equally, addressing complex computational challenges posed by the large degrees of freedom, in contrast to simpler models that focus on a subset of electronic states.

In practice, a primary challenge of \ac{QEDFT} is determining, in addition to the standard electron-electron (arising from the longitudinal Coulomb interaction), the electron-photon (transverse interaction) exchange-correlation potential for the non-interacting \ac{KS} system to reproduce the electron and photon density of the interacting and coupled system~\cite{flick.ruggenthaler.ea_2015}. 
Various perturbative approximations have emerged~\cite{pellegrini.flick.ea_2015,flick_2022} to obtain the electron-photon exchange potential, including an \ac{OEP} method~\cite{pellegrini.flick.ea_2015} and a recently-developed density-based method within first-order perturbation theory~\cite{flick_2022}, which are suitable for realistic molecules~\cite{flick.schafer.ea_2018,flick.welakuh.ea_2019}. 
Yet, these perturbative approaches lose accuracy in strongly and ultra-strongly coupled systems~\cite{flick.schafer.ea_2018}. 
To overcome these limitations, non-perturbative methods need to be developed. 
For instance, a recent method based on the photon-random-phase approximation shows promise for strong coupling in the generalized Dicke model, but its suitability for realistic systems awaits further exploration~\cite{novokreschenov.kudlis.ea_2023}.

Another non-perturbative technique~\cite{schafer.buchholz.ea_2021}, based on the local-force equation of the non-relativistic \ac{PF} Hamiltonian, approximates electron-photon exchange-correlation potentials by expressing quantum-photon fluctuations in the \ac{PF} Hamiltonian through the paramagnetic current of the matter system, simplifying the intricate photon Fock-space computations. 
This technique has been studied for a simple one-dimensional system (e.g., one-dimensional hydrogen with a soft-core Coulomb potential) and accurately reproduces the static total energy, dipole moment, and polariton spectrum, covering the whole range from weak to deep-strong light-matter coupling scenarios.
%
In its simplest form, this approximation strategy results in the \ac{px} potential and, in the homogeneous limit, gives a local-density version known as the \ac{pxLDA} functional.
Thanks to its density-based \ac{pxLDA} potential and construction in the velocity gauge, this method is adaptable for both large finite and extended systems. 
However, its effectiveness in replicating electron densities coupled to optical cavities in higher-dimensional systems remains unexplored.

In this work, we demonstrate the efficacy of the \ac{px} functional, derived from the local-force equation of the \ac{PF} Hamiltonian within the long-wavelength approximation, in accurately reproducing the electron density of one-, two-, and three-dimensional finite systems in the ultra-strong-coupling regime. 
However, as we approach the weak-coupling regime, accounting for the electron-photon correlation becomes essential to ensure accurate qualitative and quantitative electron density predictions. 
To address this, we propose the inclusion of a renormalization factor in the electron-photon functional. 
We focus on three finite one-electron systems coupled to the photon vacuum of a perfect cavity—a one-dimensional \ac{HO}, a two-dimensional quantum ring, and a three-dimensional hydrogen atom. 
All these systems are coupled to a single effective photon mode~\cite{svendsen.ruggenthaler.ea_2023} for simplicity, but it is straightforward to extend the \ac{px} potential to many photon modes due to the additive nature of the functional approximation. 
Emphasizing the electron-photon interaction, our findings provide insights into the performance of the proposed \ac{px} functional and highlight its importance in predicting light-matter interactions across different materials and scenarios.
We note that the electron-photon functional is not limited to but can be beyond the long-wavelength approximation and that the extension to the time-dependent case for driven cavities will require developing functionals depending on the time-dependent current operator, which we will show in follow-up work.

\section{Methodology}\label{sec:methodology}

\subsection{Non-relativistic Paul-Fierz Hamiltonian in the long-wavelength approximation}

We start with the non-relativistic \ac{PF} Hamiltonian $\hat{H}_{\rm{PF}}$ for $N_{e}$ electrons interacting with $M_{p}$ \textit{bare} linearly-polarized photon modes within the Coulomb gauge and in \textit{long-wavelength} approximation~\cite{svendsen.ruggenthaler.ea_2023}, i.e., the vector potential operator is $\hat{\bA}(\br) \rightarrow \hat{\bA}$. 
In Hartree atomic units, it is given as
\begin{equation}\label{eq:H_PF_start}
\begin{aligned}
    \hat{H}_{\rm{PF}}(t)=&\frac{1}{2}\sum_{l=1}^{N_{e}}\left(-i\nabla_{l}+\frac{1}{c}\hat{\bA}\right)^{2}+\frac{1}{2}\sum_{l\neq k}^{N_{e}}w(\br_{l},\br_{k}) \\ 
    & +\sum_{l=1}^{N_{e}}v_{\rm{ext}}(\br_{l},t)
     +\sum_{\alpha=1}^{M_{p}}\omega_{\alpha}\left(\hat{a}^{\dagger}_{\alpha}\hat{a}_{\alpha}+\frac{1}{2}\right).
\end{aligned}
\end{equation}
Here $l$ ($\alpha$) is the index for electrons (photon modes), $w(\br_{l},\br_{k})$ the longitudinal Coulomb interaction among electrons, $v_{\rm{ext}}(\br_{l},t)$ an external (potentially time-dependent) scalar external potential due to, e.g., the nuclei, and $\omega_{\alpha}$ and $\hat{a}_{\alpha}$ ($\hat{a}^{\dagger}_{\alpha}$) the bare photon frequency and annihilation (creation) operator for the $\alpha$-th photon mode, respectively.
The vector potential operator is
\begin{equation*}
\hat{\bA}=\sum_{\alpha=1}^{M_{p}}\hat{A}_{\alpha}\pol_{\alpha}=c\sum_{\alpha=1}^{M_{p}}\lambda_{\alpha}\pol_{\alpha}\frac{1}{\sqrt{2\omega_{\alpha}}}\left(\hat{a}^{\dagger}_{\alpha}+\hat{a}_{\alpha}\right),
\end{equation*}
where $\hat{A}_{\alpha}=(c\lambda_{\alpha}/\sqrt{2\omega_{\alpha}})(\hat{a}^{\dagger}_{\alpha}+\hat{a}_{\alpha})$, and $c$ is the speed of light and $\pol_{\alpha}$ the polarization of the $\alpha$-th bare linearly-polarized photon mode with the light-matter coupling parameter (or mode strength) $\lambda_{\alpha}$, which is proportional to the mode volume $V_{\alpha}$ via $\sqrt{1/V_{\alpha}}$ ~{\footnote{In SI units, $\lambda_{\alpha}$ is proportional to $\sqrt{\hbar/V_{\alpha}\epsilon_{0}}$ where $\hbar$ and $\epsilon_{0}$ are the reduced Plack constant and the vacuum permittivity, respectively.}}.
Note that to establish the full mapping underlying \ac{QEDFT} in the long-wavelength approximation~\cite{svendsen.ruggenthaler.ea_2023}, one also adds a mode-resolved external current as a control field for the photonic subsystem~\cite{ruggenthaler.flick.ea_2014,ruggenthaler2015ground,penz2023structure,tokatly_2013a}. 
It is straightforward to include the corresponding external and \ac{KS} currents. 
Yet, since their effects are mostly important in the time-dependent case we disregard these contributions in the following.

After the expansion of the kinetic term in Eq.~\eqref{eq:H_PF_start}, the diamagnetic term $\hat{\bA}^{2}$ can be absorbed by re-defining the bare photon modes, which become the so-called \textit{dressed} photon modes. 
The relationship between the bare and dressed photon modes can be found in Appendix~\ref{sec:appendix-dressed-photons}. 
The \ac{PF} Hamiltonian in terms of the dressed photon modes becomes
\begin{equation}\label{eq:PFH_dressed_photons}
\begin{aligned}
\hat{\tilde{H}}_{\rm{PF}}(t)=&-\frac{1}{2}\sum_{l=1}^{N_{e}}\nabla_{l}^{2}+\frac{1}{2}\sum_{l\neq k}^{N_{e}}w(\mathbf{r}_{l},\mathbf{r}_{k})+\sum_{l=1}^{N_{e}}v_{\rm{ext}}(\mathbf{r}_{l},t)
\\ & +
\frac{1}{c}\hat{\tilde{\bA}}\cdot\hat{\mathbf{J}}_{p}
+
\sum_{\alpha=1}^{M_{p}}\tilde{\omega}_{\alpha}\left(\hat{\tilde{a}}_{\alpha}^{\dagger}\hat{\tilde{a}}_{\alpha}+\frac{1}{2}\right),   
\end{aligned}
\end{equation}
where $\dac$ ($\daa$) is the creation (annihilation) operator , $\tilde{\omega}_{\alpha}$ photon frequency, $\dpola$ polarization, and $\tilde{\lambda}_{\alpha}$ light-matter coupling for the dressed photon modes. 
The vector potential operator in terms of the dressed photon modes is 
\begin{equation*}
\hat{\tilde{\bA}}=\sum_{\alpha=1}^{M_{p}}\hat{\tilde{A}}_{\alpha}\tilde{\pol}_{\alpha}=c\sum_{\alpha=1}^{M_{p}}\tilde{\lambda}_{\alpha}\dpola\frac{1}{\sqrt{2\tilde{\omega}_{\alpha}}}\left(\dac_{\alpha}+\daa_{\alpha}\right).
\end{equation*}
Here $\hat{\tilde{A}}_{\alpha}=(c\tilde{\lambda}_{\alpha}/\sqrt{2\tilde{\omega}_{\alpha}})(\hat{\tilde{a}}^{\dagger}_{\alpha}+\hat{\tilde{a}}_{\alpha})$, and 
$\hbJp=\sum_{l=1}^{N_e} (- i \nabla _{l}$) is the paramagnetic current operator. 
%

\subsection{Construction of the Kohn-Sham system in the long-wavelength approximation}

The many-body \ac{PF} Hamiltonian with the dressed photon modes [Eq.~\eqref{eq:PFH_dressed_photons}] is our starting point for constructing an auxiliary non-interacting \ac{KS} system within the \ac{QEDFT} framework~\cite{ruggenthaler.flick.ea_2014,tokatly_2013a,jestaedt.ruggenthaler.ea_2019}, which aims to reproduce the electron density (or current density if we go beyond the long-wavelength approximation and consider full minimal-coupling between light and matter) of the original interacting physical system. 
The auxiliary Hamiltonian we start with is
\begin{equation}\label{eq:HKS}
\hat{H}_{\rm{s}}(t) = \frac{1}{2}\sum_{l=1}^{N_e}\left(-i\nabla_l+\frac{1}{c}\tilde{\bA}_{\rm{s}}(t)\right)^{2} + \sum_{l=1}^{N_e}v_{\rm{s}}(\br_l,t),
\end{equation}
where $v_{\rm{s}}(\br,t)$ is an auxiliary potential and $\tilde{\bA}_{\rm{s}}(t)=\sum_{\alpha=1}^{M_{p}}\tilde{A}_{\rm{s},\alpha}(t)\tilde{\pol}_{\alpha}$ is an auxiliary classical vector potential (not an operator), constant over space, with~\cite{schafer.buchholz.ea_2021}
%
\begin{equation*}
\tilde{A}_{\rm{s},\alpha}(t) = -c\int_{-\infty}^{t}\frac{\tilde{\lambda}_{\alpha}^{2}}{\tilde{\omega}_{\alpha}}\sin\left[\tilde{\omega}_{\alpha}(t-t')\right]\tilde{\pol}_{\alpha}\cdot\bJp(t')dt',
\end{equation*}
where $\bJp(t')$ is the expectation value of the paramagnetic current operator $\hbJp$ computed with the wave function from the auxiliary Hamiltonian $\hat{H}_{\rm{s}}(t')$ at time $t'$.
This vector potential corresponds to the mean-field contribution from the transverse photon modes, and the $t \rightarrow -\infty$ can be replaced by the appropriate initial conditions that solve the mode-resolved Maxwell equation. 
We note that if we keep the (discretized) continuum of modes, we can also describe the radiative dissipation (openness) of a photonic environment from first principles~\cite{flick.welakuh.ea_2019, ruggenthaler.sidler.ea_2023}.

To define the exchange-correlation potential of \ac{KS} \ac{QEDFT}, we can use the local-force equation~\cite{tchenkoue.penz.ea_2019,ruggenthaler.tancogne-dejean.ea_2022}, which avoids the differentiability issue for energy functionals, the causality issue for action functionals in the time-dependent cases, and the numerical cost of the \ac{OEP} procedure of orbital-dependent functionals. 
%
%
The local-force equation can be obtained from the \ac{EOM} of the paramagnetic current density $\hbjp(\br)=\frac{1}{2i}\sum_{l=1}^{N_{e}}\left(\delta(\br-\br_{l})\overrightarrow{\nabla}_{l} - \overleftarrow{\nabla}_{l}\delta(\br-\br_{l})\right)$.
%
For ground-state (static) wave functions, the local-force equation for the \ac{PF} Hamiltonian is 
\begin{equation}\label{eq:lfb-PF}
    \rho(\br)\nabla v_{\rm{ext}}(\br) = \langle\hat{\mathbf{F}}_{T}(\br)\rangle_{\Psi} +  \langle\hat{\mathbf{F}}_{W}(\br)\rangle_{\Psi} - \frac{1}{c}\langle(\hat{\tilde{\bA}}\cdot\nabla)\hbjp(\br)\rangle_{\Psi}, 
\end{equation}
where $\rho(\br)$ is the electron density of the coupled light-matter ground state $\ket{\Psi}$, $\hat{\mathbf{F}}_{T}(\br)=\frac{i}{2}\left[\hbjp(\br),\sum_{l=1}^{N_{e}}\nabla^{2}_{l}\right]$ the kinetic-force density, and $\hat{\mathbf{F}}_{W}(\br)=-\frac{i}{2}\left[\hbjp(\br),\sum_{l\neq k}^{N_{e}}w(\br_{l},\br_{k})\right]$ the interaction-force density. 
Here the expectation value $\langle.\rangle$ is evaluated at the exact ground state $\ket{\Psi}$ of the \ac{PF} Hamiltonian.  
Similarly, we can find the local-force equation for the auxiliary Hamiltonian [Eq.~\eqref{eq:HKS}] 
\begin{equation}\label{eq:lfb-KS}
    \rho_{\rm{s}}(\br)\nabla v_{\rm{s}}(\br) = \langle\hat{\mathbf{F}}_{T}(\br)\rangle_{\Phi} - \frac{1}{c}(\tilde{\bA}_{\rm{s}}\cdot\nabla)\langle\hbjp(\br)\rangle_{\Phi}, 
\end{equation}
where $\ket{\Phi}$ is a Slater determinant for the ground state of the non-interacting auxiliary Hamiltonian $\hat{H}_{\rm{s}}$ and $\rho_{\rm{s}}(\br)$ is the corresponding ground-state density. 

If we now assume that both the \ac{PF} and the auxiliary Hamiltonian have the same ground-state density $\rho(\br) = \rho_{\rm{s}}(\br)$, the difference between the two local-force equations [Eqs.~\eqref{eq:lfb-PF} and~\eqref{eq:lfb-KS}] defines the \ac{Mxc} potential $v_{\rm{Mxc}}(\br)=v_{\rm{s}}(\br)-v_{\rm{ext}}(\br)$ as 
\begin{equation}\label{eq:diff-lfb-eqs}
\begin{aligned}
\rho(\br) \nabla v_{\rm{Mxc}}(\br) & =\langle\hat{\mathbf{F}}_{T}(\br)\rangle_{\Phi}
-\langle\hat{\mathbf{F}}_{T}(\mathbf{r})\rangle_{\Psi}-
\langle\hat{\mathbf{F}}_{W}(\mathbf{r})\rangle_{\Psi}
\\ & +\frac{1}{c} \langle(\hat{\tilde{\bA}}\cdot\nabla)\hbjp(\mathbf{r})\rangle_{\Psi}
-\frac{1}{c}(\tilde{\mathbf{A}}_{\rm{s}}\cdot\nabla) \langle\hbjp(\mathbf{r})\rangle_{\Phi}.
\end{aligned}
\end{equation}
For the ground-state (or static) scenarios, the constant classical vector potential $\tilde{\bA}_{\rm{s}}$ in the auxiliary Hamiltonian Eq.~\eqref{eq:HKS} can be eliminated through a gauge transformation on the ground-state wave function. 
This operation removes the last term in Eq.~\eqref{eq:diff-lfb-eqs}. 
Equation~\eqref{eq:diff-lfb-eqs} allows us to define the electron-electron and electron-photon exchange-correlation potentials.
For instance, the Hartree-exchange potential $v_{\rm{Hx}}(\br)$ for the (longitudinal) electron-electron interaction can be defined as~\cite{tchenkoue.penz.ea_2019, ruggenthaler.tancogne-dejean.ea_2022}
\begin{equation}\label{eq:vhx-def}
    \rho(\br) \nabla v_{\rm{Hx}}(\br)  = -\langle\hat{\mathbf{F}}_{W}(\mathbf{r})\rangle_{\Phi},
\end{equation}
where we replace the the exact ground-state wave function $\ket{\Psi}$ with the Slater-determinant $\ket{\Phi}$.
For the electron-photon interaction, we define the \ac{pxc} potential $v_{\rm{pxc}}(\br)$ as 
\begin{equation}\label{eq:vpxc-def}
    \rho(\br)\nabla v_{\rm{pxc}}(\br) = \frac{1}{c} \langle(\hat{\tilde{\bA}}\cdot\nabla)\hbjp(\mathbf{r})\rangle_{\Psi},
\end{equation}
where we do not know, in general, the exact wave function $\ket{\Psi}$ to obtain the \ac{pxc} potential.
Nevertheless, we can use a similar trick as for the Hatree-exchange potential to define the (transverse) electron-photon exchange potential from the light-matter interaction term $\frac{1}{c} \langle(\hat{\tilde{\bA}}\cdot\nabla)\hbjp(\mathbf{r})\rangle_{\Psi}$, together with the Breit-type approximation Eq.~\eqref{eq:Breit-ansatz} introduced in Ref.~\cite{schafer.buchholz.ea_2021} for the quantum fluctuations of the vector potential operator. This approximation for $\Delta\hat{\tilde{\bA}} = \sum_{\alpha=1}^{M_{p}}\Delta \hat{\tilde{A}}_{\alpha}\tilde{\pol}_{\alpha}$, where $\Delta \hat{O} = \hat{O}-\langle\hat{O}\rangle$, is
\begin{equation}\label{eq:Breit-ansatz}
\Delta \hat{\tilde{A}}_{\alpha}\approx -c\frac{\tilde{\lambda}_{\alpha}^{2}}{\tilde{\omega}_{\alpha}^{2}}\tilde{\pol}_{\alpha}\cdot\Delta\hat{\mathbf{J}}_{\rm{p}}.
\end{equation}
We can then construct the \ac{px} potential $v_{\rm{px}}(\br)$ from
\begin{equation*}
\begin{aligned}
& \frac{1}{c} \langle(\hat{\tilde{\bA}}\cdot\nabla)\hbjp(\mathbf{r})\rangle_{\Psi} = 
\frac{1}{c} \langle[(\langle\hat{\tilde{\bA}}\rangle_{\Psi}+\Delta\hat{\tilde{\bA}})\cdot\nabla]\hbjp(\mathbf{r})\rangle_{\Psi}  \\
& \rightarrow \frac{1}{2c} \langle[(\tilde{\bA}_{\rm{s}}+\sum_{\alpha=1}^{M_{p}}\frac{-c\tilde{\lambda}_{\alpha}^{2}}{\tilde{\omega}_{\alpha}^{2}}(\tilde{\pol}_{\alpha}\cdot\Delta\hat{\mathbf{J}}_{\rm{p}})\tilde{\pol}_{\alpha})\cdot\nabla]\hbjp(\mathbf{r})\rangle_{\Phi} +c.c. \\ 
& = -\frac{1}{2}\sum_{\alpha=1}^{M_{p}}\frac{\tilde{\lambda}_{\alpha}^{2}}{\tilde{\omega}_{\alpha}^{2}}\left[\langle(\dpola\cdot\hbJp)(\dpola\cdot\nabla)\hbjp(\mathbf{r})\rangle_{\Phi}+c.c.\right],
\end{aligned}
\end{equation*}
where we use $\hat{\tilde{\bA}}= \langle\hat{\tilde{\bA}}\rangle_{\Psi} + \Delta\hat{\bA}$. 
In the above relation we have replaced the mean-field vector potential $\langle\hat{\tilde{\bA}}\rangle_{\Psi}$ with the auxiliary classical vector potential $\tilde{\bA}_{\rm{s}}$ [Eq.~\eqref{eq:HKS}] and employed the Breit-type approximation for $\Delta\hat{\bA}$ [Eq.~\eqref{eq:Breit-ansatz}]. 
Since it is then not guaranteed anymore that this gives a real number, it is necessary to take only the real part of the expression with $c.c.$ meaning the complex conjugate.
Note that, in general, $\langle\hat{\tilde{\bA}}\rangle_{\Psi}\neq \tilde{\bA}_{\rm{s}}$, which could be controlled with an exchange-correlation current~\cite{ruggenthaler2015ground,penz2023structure}. Alternatively, we can perform a gauge transformation such that $\langle\hat{\tilde{\bA}}\rangle_{\Psi}= \tilde{\bA}_{\rm{s}}$ and we obtain in this way also the full knowledge of the photonic part.
In the last line, the classical vector potential $\tilde{\bA}_{\rm{s}}$ and the contribution from the mean-field paramagnetic current $\bJp= \langle \hbJp \rangle_\Phi$ cancel each other.
We thus define the electron-photon exchange potential $v_{\rm{px}}(\br)$ as
\begin{equation}\label{eq:vpx-def}
\rho(\br)\nabla v_{\rm{px}}(\br) = -\frac{1}{2}\sum_{\alpha=1}^{M_{p}}\frac{\tilde{\lambda}_{\alpha}^{2}}{\tilde{\omega}_{\alpha}^{2}}(\dpola\cdot\nabla)(\mathbf{f}_{\alpha,\rm{px}}(\br)+c.c.),
\end{equation}
where 
\begin{equation}\label{eq:fpx-def}
\mathbf{f}_{\alpha,\rm{px}}(\br) =\langle(\dpola\cdot\hbJp)\hbjp(\br)\rangle_{\Phi}.
\end{equation}
Next, the \ac{pc} potential $v_{\rm{pc}}(\br)$ is defined as $v_{\rm{pc}}(\br) = v_{\rm{pxc}}(\br)- v_{\rm{px}}(\br)$ and can be solved, if the exact wave function $\ket{\Psi}$ is known, using 
\begin{equation*}
\begin{aligned}
& \rho(\br)\nabla v_{\rm{pc}}(\br) = \\
& \frac{1}{c} \langle(\hat{\tilde{\bA}}\cdot\nabla)\hbjp(\mathbf{r})\rangle_{\Psi} +\frac{1}{2}\sum_{\alpha=1}^{M_{p}}\frac{\tilde{\lambda}_{\alpha}^{2}}{\tilde{\omega}_{\alpha}^{2}}(\dpola\cdot\nabla)(\mathbf{f}_{\alpha,\rm{px}}(\br)+c.c.), 
\end{aligned}
\end{equation*}
which is obtained from the difference between Eqs.~\eqref{eq:vpxc-def} and~\eqref{eq:vpx-def}.
The remaining correlation potential from both the electron-electron and electron-photon interaction, denoted as $v_{\rm{c}}(\br)$, is defined as $v_{\rm{Mxc}}(\br)-v_{\rm{Hx}}(\br)-v_{\rm{pxc}}(\br)$ and can, in principle, be obtained from Eqs.~\eqref{eq:diff-lfb-eqs},~\eqref{eq:vhx-def}, and~\eqref{eq:vpxc-def} as
\begin{equation}\label{eq:vc-def}
\begin{aligned}
&\rho(\br)\nabla v_{\rm{c}}(\br) = \\ 
&\left(\langle\hat{\mathbf{F}}_{T}(\br)\rangle_{\Phi} -\langle\hat{\mathbf{F}}_{T}(\mathbf{r})\rangle_{\Psi}\right) + \left(\langle\hat{\mathbf{F}}_{W}(\mathbf{r})\rangle_{\Phi} - \langle\hat{\mathbf{F}}_{W}(\mathbf{r})\rangle_{\Psi}\right). 
\end{aligned}
\end{equation}
%
%
%
%
Note that in principle we also have an equation of motion for the photonic part of the coupled system~\cite{ruggenthaler2015ground,penz2023structure}. However, in the static long-wavelength case this equation becomes equivalent to $\langle\hat{\tilde{\bA}}\rangle_{\Psi}= \tilde{\bA}_{\rm{s}}$. For a fixed gauge in the physical as well as auxiliary \ac{KS} system this can be achieved via an exchange-correlation current. This auxiliary current can become important to model more involved photonic observables and in the time-dependent case. Nevertheless, the photon-exchange approximation provides already access to information of the photon field as discussed Ref.~\cite{schafer.buchholz.ea_2021}. A detailed discussion of the photonic aspects is beyond the scope of this work.

After defining the exchange-correlation potentials, we notice that the formulas for the Hartree-exchange, electron-photon exchange, and correlation potentials all have the following form:
\begin{equation*}
\rho(\br)\nabla v(\br) = \mathbf{h}(\br),
\end{equation*}
where $v(\br)$ represents a potential and $\mathbf{h}(\br)$ a vector-valued function. 
The potential $v(\br)$ can be solved using the Poisson equation
\begin{equation*}
    \nabla^{2} v(\br) = \nabla\cdot\left(\frac{\mathbf{h}(\br)}{\rho(\br)}\right).
\end{equation*}
This approach has been implemented in the \textit{Octopus} code~\cite{tancogne-dejean.oliveira.ea_2020} to obtain the Hartree-exchange potential~\cite{ruggenthaler.tancogne-dejean.ea_2022}. 
Similarly, the \ac{px} potential $v_{\rm{px}}(\br)$ can be obtained by solving the following Poisson equation, which is derived from Eq.~\eqref{eq:vpx-def}, 
\begin{equation} \label{eq:vpx-poisson}
\nabla^{2}v_{{\rm{px}}}(\br) = -\nabla\cdot\left[\sum_{\alpha=1}^{M_{p}}\frac{\tilde{\lambda}_{\alpha}^{2}}{2\tilde{\omega}_{\alpha}^{2}}\frac{(\dpola\cdot\nabla) \left(\mathbf{f}_{\alpha,\rm{px}}(\br)+c.c.\right)}{\rho(\br)}\right].
\end{equation}
%
%
%
%
%
For one-electron systems coupled to one single dressed photon mode with the frequency $\tilde{\omega}$, light-matter coupling $\tilde{\lambda}$, and polarization direction $\tilde{\pol}$,  the \ac{px} potential $v_{{\rm{px}}}(\br)$ can be obtained directly from the electron density (for details see Appendix~\ref{sec:appendix-1pxc}) using 
\begin{equation}\label{eq:vpx-one-electron}
v_{\rm{px}}(\mathbf{r}) = \frac{{\tilde{\lambda}}^{2}}{2\tilde{\omega}^{2}} \frac{(\tilde{\pol}\cdot\nabla)^{2}\rho^{\frac{1}{2}}(\br)}{\rho^{\frac{1}{2}}(\br)}.
\end{equation}
In the homogeneous density limit, which leads to the \ac{LDA}, the expectation value in Eq.~\eqref{eq:fpx-def} can be evaluated in terms of a Slater determinant of plane waves and leads to~\cite{schafer.buchholz.ea_2021}
\begin{equation}\label{eq:sub-pxforce}
     \mathbf{f}_{\alpha,\rm{px}}(\br) \rightarrow \mathbf{f}_{\alpha,\rm{pxLDA}}(\br) = \frac{2 V_{d}}{(2\pi)^{d}}\frac{k_{\rm{F}}^{d+2}(\br)}{d+2}\dpola,
\end{equation}
where $k_{\rm{F}}(\br)=2\pi(\rho(\br)/2V_{d})^{1/d}$ and $V_{d}$ is the volume of the $d$-dimensional unit sphere (i.e., $V_{1} =2$, $V_{2} = \pi$, and $V_{3}=4\pi/3$).
It can be shown that the force $\mathbf{f}_{\alpha,\rm{px}}(\br)$ satisfies the zero-force condition [see Eq.~\eqref{eq:zero-pxlda-force} in Appendix~\ref{sec:epxlda}].
The \ac{px} potential within the homogeneous limit then becomes an explicit density functional and can be calculated by solving 
\begin{equation}\label{eq:pxlda-d-dimension}
\nabla^{2} v_{{\rm{pxLDA}}}(\br) = -\sum_{\alpha=1}^{M_{p}}\frac{2 \pi^{2}\tilde{\lambda}_{\alpha}^{2}}{\tilde{\omega}_{\alpha}^{2}}\left[(\dpola\cdot\nabla)^{2}\left(\frac{\rho(\br)}{2V_{d}}\right)^{\frac{2}{d}}\right].
\end{equation}
In one dimension and for isotropic problems, the \ac{pxLDA} potential has an explicit form respectively,
\begin{align*}
    & v_{\rm{pxLDA}}(x)=-\frac{\pi^{2}}{8} \sum_{\alpha=1}^{M_{p}} \frac{\tilde{\lambda}_{\alpha}^{2}}{\tilde{\omega}_{\alpha}^{2}} \rho^{2}(x), \\
    & v_{\rm{pxLDA}}^{\rm{iso}}(\br) = -\frac{2\pi^{2}}{d} \sum_{\alpha=1}^{M_{p}}\frac{\tilde{\lambda}_{\alpha}^{2}}{\tilde{\omega}_{\alpha}^{2}} \left(\frac{\rho(\br)}{2V_{d}}\right)^{\frac{2}{d}}.
\end{align*}
For two and three dimensions one needs to solve the Poisson equation [Eq.~\eqref{eq:pxlda-d-dimension}], using either the conjugate gradient or the Poisson-kernel method~\cite{tancogne-dejean.oliveira.ea_2020}.
%
%
%
Yet, the correlation potential $v_{\rm{c}}(\br)$ remains unknown in general, as it depends on the exact ground-state wave function [Eq.~\eqref{eq:vc-def}], necessitating the use of alternative numerical methods like quantum Monte Carlo~\cite{weber.vinasbostrom.ea_2023,weber2023eps}.
Nevertheless, we introduce a method in Sec.~\ref{sec:perturbation-analysis} to explore one aspect of the correlation potential, namely, the electron-photon correlation potential, in the context of weak coupling in the light-matter interaction. 
%

%
%
After obtaining the \ac{px} potential from the local-force equation, we explore the associated \ac{px} energy below.
We note that when employing Eq.~\eqref{eq:Breit-ansatz} in the \ac{PF} Hamiltonian [Eq.~\eqref{eq:PFH_dressed_photons}], we need to consistently also take the contribution of the photonic energy into account~\cite{schafer.buchholz.ea_2021}. That is, the replacement of the photonic operator with the paramagnetic current operator in the last term of Eq.~\eqref{eq:PFH_dressed_photons} gives a counteracting contribution to the light-matter interaction term $\hat{\tilde{\mathbf{A}}}\cdot\hbJp$. 
%
A comparison with the \ac{EOM} of a correspondingly defined Breit-type Hamiltonian~\cite{schafer.buchholz.ea_2021} reveals that the substitution leads to 
%
%
\begin{equation*}
\frac{1}{c} \langle \hat{\tilde{\mathbf{A}}}\cdot\hbJp \rangle_{\Psi} + \sum_{\alpha=1}^{M_p}\tilde{\omega}_{\alpha} \langle \hat{\tilde{a}}_{\alpha}^{\dagger} \hat{\tilde{a}}_{\alpha}\rangle_{\Psi} \rightarrow  E_{\rm px}[\rho],
\end{equation*}
%
where the electron-photon exchange energy is defined as
\begin{equation}\label{eq:def-pxenergy-functional}
E_{\rm{px}}[\rho] = -\sum_{\alpha=1}^{M_{p}}\frac{\tilde{\lambda}_{\alpha}^{2}}{2\tilde{\omega}_{\alpha}^{2}}\braket{(\tilde{\pol}_{\alpha}\cdot\hbJp)\Phi[\rho]}{(\tilde{\pol}_{\alpha}\cdot\hbJp)\Phi[\rho]}.
\end{equation}
Here the factor of $1/2$ results from the counteracting photonic energy contribution.
In the homogeneous limit, the \ac{px} energy becomes (for details see Appendix~\ref{sec:epxlda})
\begin{equation}\label{eq:epxlda-formula}
E_{\rm pxLDA}[\rho]= \frac{-2\pi^{2}}{(d+2)({2V_{d}})^{\frac{2}{d}}}\sum_{\alpha=1}^{M_{p}}\frac{\tilde{\lambda}_{\alpha}^{2}}{\tilde{\omega}_{\alpha}^{2}}\int d\br\ \rho^{\frac{2+d}{d}}(\br).
\end{equation}
This form can be derived either through applying the \ac{LDA} on the energy functional of Eq.~\eqref{eq:def-pxenergy-functional} or via the exchange virial relation using the \ac{LDA} for the force from Eq.~\eqref{eq:sub-pxforce} (see Appendix~\ref{sec:epxlda} for details).
We note that the exchange force, which in general has transverse components, needs to be taken into account to fulfill the exchange virial relation~\cite{ruggenthaler.tancogne-dejean.ea_2022}. 
%
%
When determining the \ac{pxLDA} potential from the functional derivative of the \ac{pxLDA} energy [Eq.~\eqref{eq:epxlda-formula}], we end up with the isotropic \ac{pxLDA} potential $v^{\rm{iso}}_{\rm{pxLDA}}(\br)$. 
However, this isotropic potential lacks information about the polarization of photon modes. 
To preserve this information, it is essential to use the \ac{pxLDA} potential obtained from the local-force equation in the \ac{KS} equations and subsequently compute the \ac{pxLDA} energy after obtaining the electron density.

%
%
In practice, the \ac{KS} Hamiltonian, which is designed to reproduce the electron density of the \ac{PF} Hamiltonian [Eq.~\eqref{eq:PFH_dressed_photons}], has to be solved in a self-consistent way for ground-state calculations similar to standard \ac{DFT}~\cite{dreizler2012density,martin_2020}:
\begin{enumerate}
\item Calculate the \ac{Mxc} potential $v_{\rm{Mxc}}(\br)$ using either the \ac{KS} orbitals or the electron density;
\item Construct the \ac{KS} Hamiltonian $\hat{H}_{\rm{KS}}$ using the Hamiltonian of Eq.~\eqref{eq:HKS} with $v_{\rm KS}(\br) = v_{\rm ext}(\br) + v_{\rm Mxc}(\br)$ and without the vector potential in the time-independent cases;
%
\item Solve the resulting \ac{KS} Hamiltonian and obtain the \ac{KS} orbitals and electron density, which are used in Step 1 to get the associated \ac{Mxc} potentials;
\item Loop through Step 1 to 3 until the electron density converges within a desired threshold. 
\end{enumerate}
%
%
For time-dependent calculations, once the ground state of the \ac{KS} Hamiltonian $\ket{\Phi_{\rm{KS}}}$ is obtained, the time propagation of the ground state is determined by solving the non-linear Schrödinger-type evolution equation $i\partial_{t} \ket{\Phi_{\rm{KS}}(t)} = \hat{H}_{\rm{KS}}(t)\ket{\Phi_{\rm{KS}}(t)} $ with the time-dependent Hamiltonian from Eq.~\eqref{eq:HKS}, together with the auxiliary classical vector potential $\tilde{\bA}_{\rm{s}}(t)$, and replacing $v_{\rm{s}}(\br,t) \rightarrow v_{\rm KS}(\br,t)$~\cite{ullrich2011time,ruggenthaler.penz.ea_2015,ruggenthaler.flick.ea_2014, flick.ruggenthaler.ea_2015}. 
Note that the \ac{Mxc} potential based on the local-force equation of Eq.~\eqref{eq:diff-lfb-eqs} is strictly speaking \textit{only} for the static case, making it an adiabatic approximation when used in time-dependent simulations.
Although it is possible to derive non-adiabatic potentials (see Refs.~\cite{tchenkoue.penz.ea_2019,schafer.buchholz.ea_2021, ruggenthaler.tancogne-dejean.ea_2022} for details), this is beyond the scope of our paper. 
Below we thus use the adiabatic approximation to obtain associated spectra~\cite{ullrich2011time, ruggenthaler.penz.ea_2015}.

%
%

\subsection{The weak-coupling limit: perturbation-theory analysis}\label{sec:perturbation-analysis}

The \ac{px} potential has been derived from the \ac{PF} Hamiltonian using the photon-coupled homogeneous electron-gas basis~\cite{schafer.buchholz.ea_2021, rokaj.ruggenthaler.ea_2022d}. 
In the limit $\lambda_{\alpha} \rightarrow \infty$ and $\omega_{\alpha} \rightarrow \infty$, the \ac{px} potential becomes the sole contribution. 
%
While it vanishes for $\lambda_{\alpha} \rightarrow 0$ (as it should), its behavior in this limit has not been extensively studied. 
For simplicity, we focus on one effective mode~\cite{svendsen.ruggenthaler.ea_2023}, as simulations with numerous modes pose numerical challenges for exact reference calculations. 
This is the focus of this section.
However, the explored approximate electron-photon functionals can be easily extended to accommodate any number of photon modes without significant numerical overhead.


To better understand the applicability of the \ac{px} and \ac{pxLDA} approximation for the \ac{PF} Hamiltonian, we compare it with static perturbation theory. 
%
%
Our starting Hamiltonian with one dressed photon mode, i.e., Eq.~\eqref{eq:PFH_dressed_photons} with one mode, is rewritten as 
\begin{equation}\label{eq:PF-one-mode}
    \hat{\tilde{H}}_{\rm{PF}} = \hat{H}_{\rm{M}} + \frac{1}{c}\hat{\tilde{\bA}}\cdot\hbJp + \hat{H}_{\gamma},
\end{equation}
where the Hamiltonian for the matter subsystem is
\begin{equation}\label{eq:matter-hamiltonian}
    \hat{H}_{\rm{M}} = -\frac{1}{2}\sum_{l=1}^{N_{e}}\nabla_{l}^{2}+\sum_{l=1}^{N_{e}}v_{\rm{ext}}(\mathbf{r}_{l})+\frac{1}{2}\sum_{l\neq k}^{N_{e}}w(\mathbf{r}_{l},\mathbf{r}_{k}),
\end{equation}
and the Hamiltonian for the dressed photon mode is 
\begin{equation*}
    \hat{H}_{\gamma}=\tilde{\omega}\left(\dac\daa+\frac{1}{2}\right).
\end{equation*} 
In this section we overload the notation $\hat{\tilde{H}}_{\rm{PF}}$ for one photon mode~\cite{svendsen.ruggenthaler.ea_2023}, compared to Eq.~\eqref{eq:PFH_dressed_photons}. If we mention one-mode cases, we refer to Eq.~\eqref{eq:PF-one-mode}; otherwise, we refer to the more general multi-mode form of Eq.~\eqref{eq:PFH_dressed_photons}.

We denote $\hat{H}_{\rm{M}}\ket{m^{(0)}}={\epsilon}_{m}^{(0)}\ket{m^{(0)}}$, where ${\epsilon}_{m}^{(0)}$ is the energy for the $m$-th unperturbed many-body matter state $\ket{m^{(0)}}$.
Furthermore, we have $\hat{H}_{\gamma}\ket{\tilde{n}}=\tilde{\omega}(\tilde{n}+1/2)\ket{\tilde{n}}$, where $\tilde{\omega}$ and $\tilde{n}$ are the dressed photon frequency and photon number for the photon mode $\ket{\tilde{n}}$, respectively.
The vector potential for the dressed cavity is $\hat{\tilde{\bA}}=\hat{\tilde{A}}\tilde{\pol}=(c\tilde{\lambda}/\sqrt{2\tilde{\omega}})(\dac+\daa)$, where $\tilde{\lambda} = \lambda$ and $\tilde{\omega}^{2} = \omega^{2} + N_{e}\lambda^{2}$ (see Appendix~\ref{sec:appendix-dressed-photons}).
%
%
Next, we assume weak light-matter coupling such that the light-matter interaction can be considered as a perturbation to the matter Hamiltonian [Eq.~\eqref{eq:matter-hamiltonian}],
\begin{equation*}
    \Delta \hat{V} = \frac{1}{c}\hat{\tilde{\bA}}\cdot\hbJp = \frac{\tilde{\lambda}}{\sqrt{2\tilde{\omega}}}(\dac+\daa)(\tilde{\pol}\cdot\hbJp).
\end{equation*}
The unperturbed system is the composite system that consists of the ground state of the matter $\ket{0^{(0)}}$ and that of the one-dressed-photon-mode subsystem $\ket{\tilde{0}}$. 
We write the ground state of the unperturbed composite system as a direct tensor product state $|{\Psi_{0}^{(0)}}\rangle = \ket{0\tilde{0}} = \ket{0^{(0)}}\otimes\ket{\tilde{0}}$.

In the weak light-matter coupling regime, the modified ground-state wave function $\ket{\Psi_{0}}$ up to the 1st-order correction is 
\begin{equation*}
    \ket{\Psi_{0}} \approx |{\Psi_{0}^{(0)}}\rangle +|{\Psi_{0}^{(1)}}\rangle,
\end{equation*}
where $|{\Psi_{0}^{(1)}}\rangle$ is the 1st-order correction to the ground-state wave function, which only involves the contribution from the $1$st photon sector $\ket{\tilde{1}}$,
\begin{equation*}
|{\Psi_{0}^{(1)}}\rangle = -\frac{\tilde{\lambda}}{\sqrt{2}\tilde{\omega}^{3/2}} \sum_{m=0}^{\infty}\frac{\mel{m^{(0)}}{(\tilde{\pol}\cdot\hbJp)}{0^{(0)}}}{1+\Delta \epsilon_{m0}^{(0)}/\tilde{\omega}} \ket{m\tilde{1}},
\end{equation*}
where $\Delta \epsilon_{m0}^{(0)} = \epsilon_{m}^{(0)}-\epsilon_{0}^{(0)}$ and $\ket{m\tilde{1}} = \ket{m^{(0)}}\otimes\ket{\tilde{1}}$.
The modified ground-state wave function is an entangled state, which cannot be rewritten as a single tensor product state of the matter and photonic state.

Using the modified ground-state wave function $\ket{\Psi_{0}}$ and Eq.~\eqref{eq:vpxc-def}, we evaluate the \ac{pxc} potential up to the lowest order of the light-matter coupling, 
\begin{equation*}
\begin{aligned}
& \rho(\br)\nabla v_{\rm{pxc}}(\br) \approx \frac{1}{c} \langle \Psi_{0}^{(1)} | (\hat{\tilde{\bA}}\cdot\nabla)\hbjp(\mathbf{r}) | \Psi_{0}^{(0)} \rangle + c.c. \\
& = -\frac{\tilde{\lambda}^{2}}{2\tilde{\omega}^{2}}\sum_{m=0}^{\infty}
\frac{\langle 0^{(0)}|(\tilde{\pol}\cdot\hbJp)|m^{(0)}\rangle \langle m^{(0)}|(\tilde{\pol}\cdot\nabla)\hbjp(\mathbf{r})|0^{(0)}\rangle}{1+\Delta \epsilon_{m0}^{(0)}/\tilde{\omega}} \\ 
& \ \ \ \ + c.c. \\ 
& = -\frac{\tilde{\lambda}^{2}}{2\tilde{\omega}^{2}}\langle(\tilde{\pol}\cdot\hbJp)(\tilde{\pol}\cdot\nabla)\hbjp(\br)\rangle_{\Psi_{0}^{(0)}} \\
& \ \ \ \ +\frac{\tilde{\lambda}^{2}}{2\tilde{\omega}^{2}}\sum_{m=1}^{\infty}
\frac{\langle 0^{(0)}|(\tilde{\pol}\cdot\hbJp)|m^{(0)}\rangle \langle m^{(0)}|(\tilde{\pol}\cdot\nabla)\hbjp(\mathbf{r})|0^{(0)}\rangle}{1+\tilde{\omega}/\Delta \epsilon_{m0}^{(0)}} \\
& \ \ \ \ + c.c., 
\end{aligned}    
\end{equation*}
where we have only the contribution between $| \Psi_{0}^{(0)} \rangle$ and $ | \Psi_{0}^{(1)} \rangle$ due to the orthogonality of the photonic states. In the last line we separate a \ac{px}-like contribution, where we use $|\Psi_{0}^{(0)}\rangle$ instead of $|\Phi\rangle$ in Eq.~\eqref{eq:fpx-def}, from the \ac{pxc} potential with the help of the identity $\mathbbm{1} = \sum_{m=0}^\infty\ket{m^{(0)}}\bra{m^{(0)}}$.
Then, the \ac{pc} potential in the weak-coupling regime is
\begin{equation}\label{eq:vpc-perturbation}
\begin{aligned}
& \rho(\br)\nabla v_{\rm{pc}}(\br) \approx \\
& \frac{\tilde{\lambda}^{2}}{2\tilde{\omega}^{2}}\sum_{m=1}^{\infty}
\frac{\langle 0^{(0)}|(\tilde{\pol}\cdot\hbJp)|m^{(0)}\rangle \langle m^{(0)}|(\tilde{\pol}\cdot\nabla)\hbjp(\mathbf{r})|0^{(0)}\rangle}{1+\tilde{\omega}/\Delta \epsilon_{m0}^{(0)}} + c.c.
\end{aligned}    
\end{equation}
When $\tilde{\omega}\rightarrow \infty$ (i.e., in the strong-coupling regime or large photon frequency regime), the \ac{pc} potential vanishes, and the \ac{pxc} becomes a \ac{px}-like potential, consistent with the results obtained from the analysis of the photon-coupled homogeneous electron-gas basis~\cite{schafer.buchholz.ea_2021}.

Computing the \ac{pc} potential in the weak coupling [Eq.~\eqref{eq:vpc-perturbation}] requires all the information of the many-body matter states, which poses numerical challenges. 
Alternatively, the light-matter interaction in the weak-coupling regime can be accurately obtained using the \ac{OEP} approach~\cite{flick.schafer.ea_2018,flick.welakuh.ea_2019}. 
Instead of solving Eq.~\eqref{eq:vpc-perturbation}, we here propose the following formula with a correlation factor $\xi_{\rm{c}}$ to approximate the \ac{pc} potential in the weak-coupling regime for one photon mode~\footnote{For multi-mode cases, each photon mode has its correlation factor $\xi_{c,\alpha}$. The correlation factor $\xi_{{\rm{c}},\alpha} = \xi_{{\rm{c}},\alpha}(\{\lambda_{\alpha}\},\{\omega_{\alpha}\})$ for each photon mode $\alpha$ is a function of the light-matter coupling as well as the photon frequency.}, 
\begin{equation*}
\rho(\br)\nabla v_{\rm{pc}}(\br) \approx \frac{\xi_{c}\tilde{\lambda}^{2}}{2\tilde{\omega}^{2}}\langle(\tilde{\pol}\cdot\hbJp)(\tilde{\pol}\cdot\nabla)\hbjp(\br)\rangle_{\Psi_{0}^{(0)}} + c.c.
\end{equation*}
This formula is obtained by assuming $\Delta \epsilon^{(0)}_{m0} \gg \tilde{\omega}$ in the second line of Eq.~\eqref{eq:vpc-perturbation} such that the dependence of the denominator on the index $m$ can be neglected.
The (positive) correlation factor $\xi_{\rm{c}}$ depends on the light-matter coupling and photon frequency, and vanishes when $\tilde{\omega}\rightarrow \infty$, i.e., in the strong-coupling or large photon frequency regime.
Thus, we use the following formula for the \ac{pxc} potential when exploring the weak-coupling regime, 
\begin{equation*}
\rho(\br)\nabla v_{\rm{pxc}}(\br) \approx -\frac{\eta_{\rm{c}}\tilde{\lambda}^{2}}{2\tilde{\omega}^{2}}\langle(\tilde{\pol}\cdot\hbJp)(\tilde{\pol}\cdot\nabla)\hbjp(\br)\rangle_{\Phi} + c.c.,
\end{equation*}
where we replace $|\Psi_{0}^{(0)}\rangle$ with $|\Phi\rangle$ and use Eq.~\eqref{eq:vpx-def}, and we define a renormalization factor $\eta_{\rm{c}}=1-\xi_{\rm{c}}$ for the \ac{px} potential to take the electron-photon correlation contribution into account. 
We denote this formula as the $\eta_{\rm{c}}$-\ac{px} approach here and below.
The renormalization factor $\eta_{\rm{c}}$ is determined by perturbation theory, e.g., comparing the results with the exact or \ac{OEP} approach.
Hereafter, we apply the approximated \ac{pxc} potential to various one-electron systems. It is anticipated that the outcomes may differ from the exact results, as the correlation potential from the kinetic- and interaction-force density [Eq.~\eqref{eq:vc-def}], and the electron-photon correlation potential arising from employing the Slater determinant in the \ac{pxc} potential instead of the exact wave function, are neglected.

\section{Results and discussion}

We analyze the performance of the different functionals derived in the previous section by applying them to three different one-electron systems coupled to an effective single mode of a perfect cavity, e.g., Fabry-Perot cavity~\footnote{For single-cavity-mode cases, the \ac{px} and \ac{pc} potentials depend only on the ratio between the light-matter coupling and bare photon frequency, i.e., $\lambda_{\alpha}/\omega_{\alpha}$}. 
Specifically, we consider electron-photon interaction in a one-dimensional \ac{HO}, a two-dimensional quantum ring, and a three-dimensional hydrogen atom. 
%
%
%
In one- and two-dimensional systems, we can directly solve the corresponding \ac{PF} Hamiltonians using exact diagonalization, which serves as a benchmark for approximate \ac{QEDFT} results.
We focus the analysis and comparison on the electron density, since this is the fundamental quantity of \ac{QEDFT} in the long-wavelength limit. 
%
For the three-dimensional hydrogen atom, we examine the ground-state electron density by comparing the \ac{OEP} approach to the \ac{px} and \ac{pxLDA} approximations. 
Additionally, for the hydrogen atom, we explore its optical absorption spectra through time-dependent \ac{QEDFT}~\cite{tokatly_2013a, ruggenthaler.flick.ea_2014,flick.ruggenthaler.ea_2015} (see Appendix~\ref{sec:comp-details} for all computational details).
%
%
Though our main focus is on investigating the behavior of electron-photon interaction without complicating the study with (approximate) electron-electron interaction, in Appendix~\ref{sec:He-example} we also present a He atom to showcase that the approach is directly applicable to multi-electron problems.

\subsection{Harmonic oscillator coupled to a photon mode}

\begin{figure}[t]
\centering
\includegraphics[width=\linewidth]{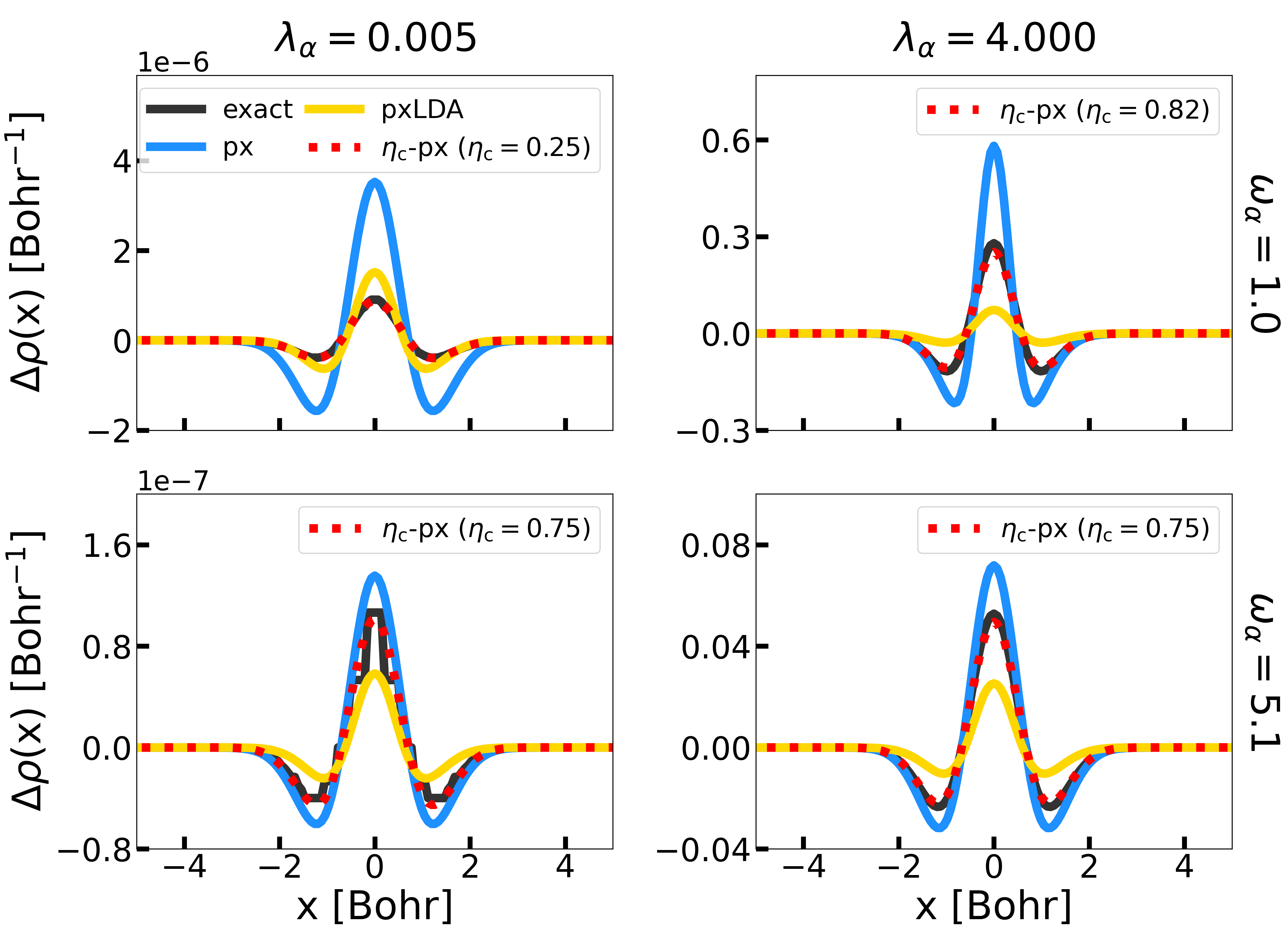}
\caption{
%
%
Electron-density differences between the inside and outside cavity, $\Delta\rho(x) = \rho_{\lambda}(x)-\rho_{\lambda=0}(x)$, in the weak- and strong-coupling regimes. 
The results are computed via the exact diagonalization and different approximate \ac{QEDFT} functionals.
}
\label{fig01}
\end{figure}

We first consider a one-dimensional \ac{HO} with the external potential $v_{\rm{SHO}}(x) = x^{2}/2$ coupled to a photon mode (see the computational details in Appendix~\ref{sec:comp-details}).
%
%
We explore both weak- and strong-coupling regimes by using two different light-matter couplings, $\lambda = 0.005$ and $4.0$, along with two photon frequencies, $\omega= 1.0$ and $5.1$.
%
%
Figure~\ref{fig01} shows the electron density differences between inside and outside the cavity, $\Delta\rho(x) = \rho_{\lambda}(x)-\rho_{\lambda=0}(x)$. 
In the weak-coupling ($\lambda = 0.005$) and low-photon-frequency ($\omega=1.0$, i.e., on resonance with the $1$st excited state of the \ac{HO}) 
scenario, both \ac{px} and \ac{pxLDA} approximations overestimate the effect of the mode on the  electron density. 
Following the discussion in Sec.~\ref{sec:perturbation-analysis}, we can rectify this by introducing the perturbation-based renormalization factor ($\eta_{\rm{c}} = 1/4$) for the \ac{px} potential.
%
%
%
In the weak-coupling but high-photon-frequency ($\omega_{\alpha}=5.1$) case, the \ac{px} functional slightly overestimates the electron density, while the \ac{pxLDA} approximation slightly underestimates it. 
The $\eta_{\rm c}$-\ac{px} functional reproduces the exact result, and the $\eta_{\rm{c}}$ factor becomes close to 1.0.

In the strong-coupling regime ($\lambda = 4.0$ and $\lambda/\omega \ge 0.1$) both, the \ac{px} and \ac{pxLDA} approximations, exhibit small deviations from the exact electron density, similar to the scenario seen in the weak coupling but with a large photon frequency ($\lambda = 0.005$ and $\omega =5.1$). 
The $\eta_{\rm c}$-\ac{px} functional can be used to restore the exact results, and the $\eta_{\rm{c}}$ factor approaches $1.0$ in this context, implying that the electron-photon correlation becomes small. 
These numerical findings demonstrate yet again that both the \ac{px} and \ac{pxLDA} approximate functionals perform well in the strong-coupling regime. 
While the \ac{pxLDA} functional tends to underestimate the effect of the cavity on the electron density compared to the \ac{px} approximation, it captures the changes qualitatively correct for one-dimensional problems. 
The presented approximation strategy also gives an intuitive understanding of the effect of the cavity in the strong-coupling regime. 
Expressing the photon fluctuation operator via the current operator [Eq.~\eqref{eq:Breit-ansatz}] leads to a renormalization of the electron mass (along the polarization direction), a feature that is also predicted for the \ac{HEG} coupled to an optical cavity~\cite{rokaj.ruggenthaler.ea_2022d}. Since this increases the effective mass of the electron, i.e., the kinetic contribution in the Schr\"odinger equation is decreased, the electron becomes more localized.
%

\subsection{Quantum ring coupled to a photon mode}

\begin{figure}[!t]
\centering
\includegraphics[width=\linewidth]{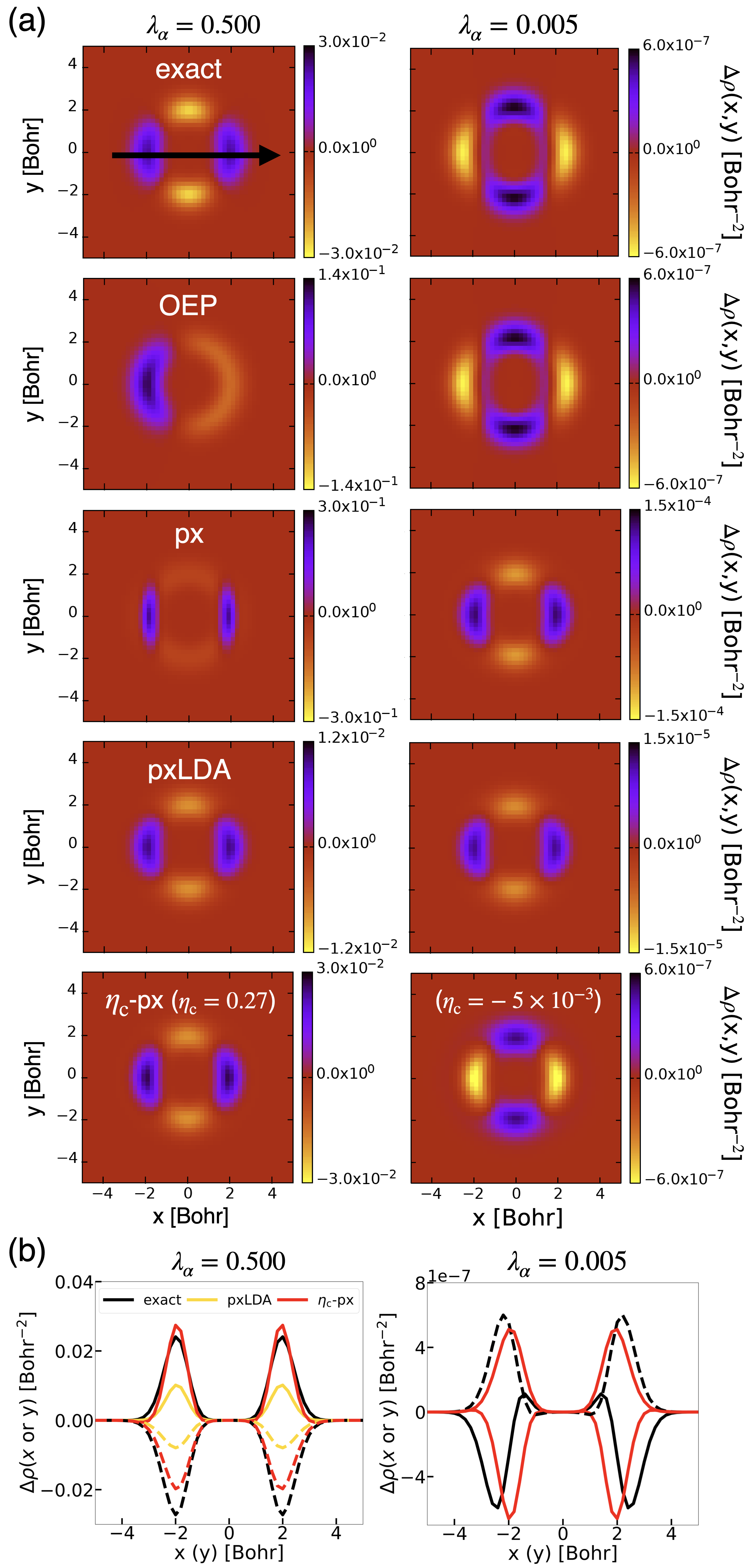}
\caption{
%
%
The photon mode is polarized along the $x$ direction (the solid right arrow) and has a frequency of $\omega = 0.125$, in resonance with the $1$st excited state of the two-dimensional quantum ring. 
a) The electron-density difference inside and outside the cavity for the exact and various approximate \ac{QEDFT} functionals in both strong- and weak-coupling regimes.  
b) The electron-density difference along the $x$ (solid lines) and $y$ (dashed lines) directions (the cut through the center of the quantum ring) in both strong- and weak-coupling scenarios. 
The renormalization factor $\eta_{\rm{c}}$ are $0.27$ and $-5\times10^{-3}$ in the strong- and weak-coupling regime, respectively.
}
\label{fig02}
\end{figure}

When the photon mode is in resonance with the $1$st excited state of a quantum ring ($\Delta \epsilon_{10}^{(0)} = 0.125$ for a confining potential $v_{\rm{QR}}(\br) = \xi_{1}|\br|^{2}/2+\xi_{2}\exp\left(-|\br|^{2}/\xi_{3}^{2}\right)$, with $\xi_{1}=0.7827$, $\xi_{2}=17.7$, and $\xi_{3}=0.997$~\cite{schafer.ruggenthaler.ea_2018}) in the weak-coupling regime, the electron density accumulates perpendicular to the polarization direction, which is set to $x$ here, while in the strong-coupling regime, the density aligns along the polarization direction [see Fig.~\ref{fig02}(a)] (for the computational details see Appendix~\ref{sec:comp-details}).
This shift in behavior makes the quantum ring an excellent case to test if the approximate \ac{QEDFT} functionals can reproduce the correct feature of the electron density from the weak- to strong-coupling regimes.
%
%
%

Figure~\ref{fig02}(a) shows electron density differences within and outside the cavity under both strong- and weak-coupling conditions, considering the different approximate \ac{QEDFT} functionals.
In the strong-coupling regime ($\lambda = 0.5$), the \ac{OEP} functional in exchange-approximation fails to qualitatively replicate the exact electron density~\cite{flick.schafer.ea_2018}, whereas the \ac{px} functional, while providing a qualitative match, tends to overestimate the effect of the cavity. 
The \ac{pxLDA} functional yields an improved agreement with the exact result in the strong-coupling regime. 
Additionally, we see that the $\eta_{\rm c}$-\ac{px} functional helps to recover quantitatively the exact results for this case. 
Figure~\ref{fig02}(b) provides a comparison of electron-density differences along the $x$ and $y$ directions for the exact, \ac{pxLDA}, and $\eta_{\rm c}$-\ac{px} methods.
In the strong-coupling regime, the electron accumulates and behaves more classically along the polarization, similar to the one-dimensional \ac{HO} case. 

In the weak-coupling regime ($\lambda = 0.005$), neither the \ac{px} nor the \ac{pxLDA} approximations capture the desired feature [Fig.\ref{fig02}(a)].
In contrast, the $\eta_{\rm c}$-\ac{px} approach manages to capture the weak-coupling feature, even though small deviations still exist [Fig.~\ref{fig02}(b)] due to the approximation we made for the $\eta_{\rm c}$-\ac{px} potential developed in Sec.~\ref{sec:perturbation-analysis} and the neglect of other correlation potentials.
%
%
%
These observations highlight the electron-photon correlation contribution in the weak-coupling regime. 
Conversely, in the strong-coupling regime, we can use approximate \ac{px} potentials alone to capture the strong-coupling features because the \ac{pc} potential quickly vanishes with $\tilde{\omega}_{\alpha}$ as discussed in Sec.~\ref{sec:perturbation-analysis}. 

\subsection{Hydrogen atom coupled to a photon mode}

We now turn our attention to a more realistic system, a hydrogen atom coupled to a photon mode (see Appendix~\ref{sec:comp-details} for the computational details).
%
%
While this system might seem simple with one atom and one electron coupled to one photon mode, the exact diagonalization of the \ac{PF} Hamiltonian in real space to obtain the electron density poses already computational challenges, demanding significant memory resources to converge the results with respect to the real-space grid size and the number of photon Fock states.
As an alternative, we rely on the results obtained from the \ac{OEP} functional in exchange-approximation as our reference, especially, in the weak-coupling regime, where the \ac{OEP} approach works well.

\begin{figure}[t]
\centering
\includegraphics[width=\linewidth]{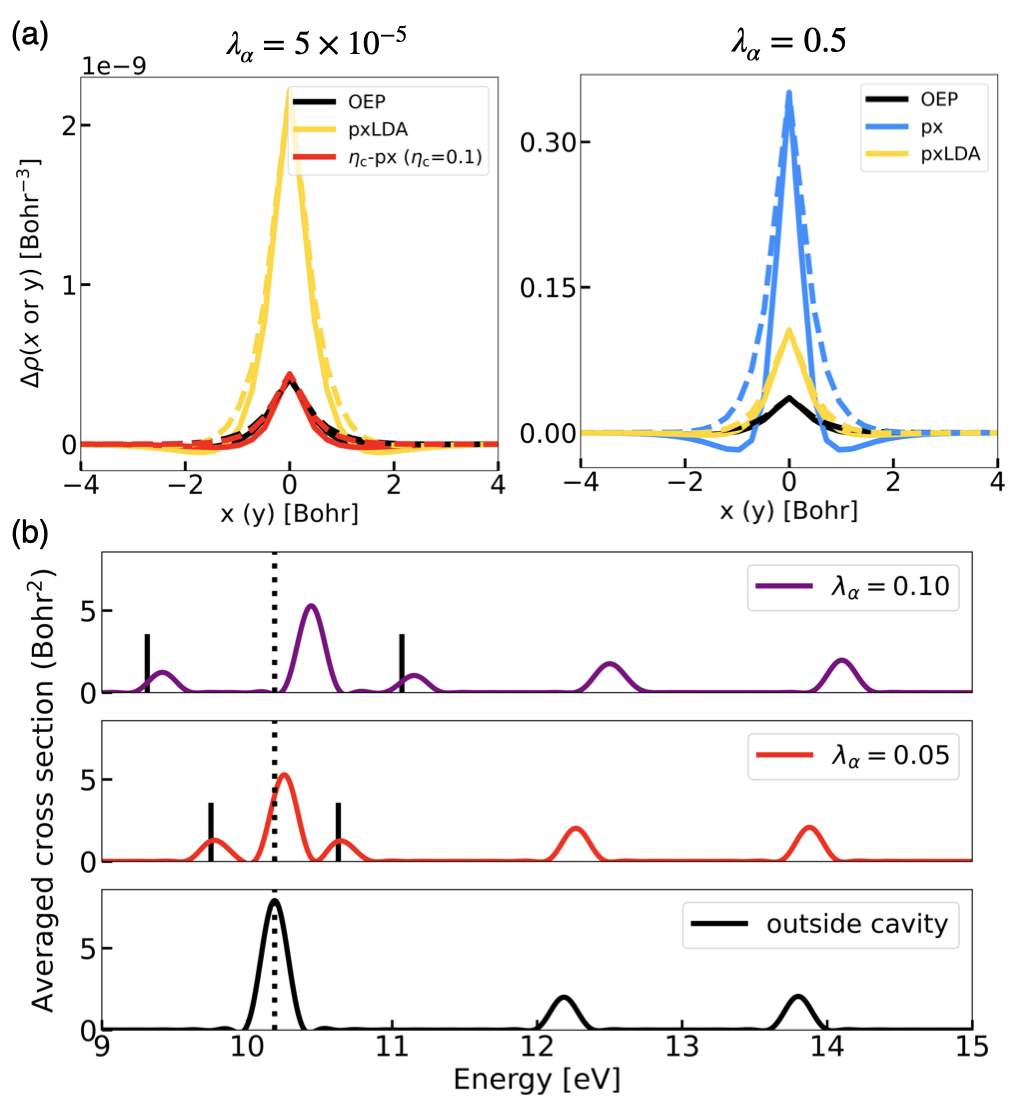}
\caption{
%
%
The photon mode is polarized along the $x$ direction and a photon frequency of $0.3745$ Hartree ($10.19$ eV), in resonance with the $1$st excited state of the hydrogen atom.
a) The electron-density differences $\Delta\rho$ along the x (solid lines) and y (dashed lines) directions in the weak- and strong-coupling regimes, computed using various \ac{QEDFT} approximations. 
b) The optical spectrum of the hydrogen atom inside and outside of the cavity. 
The results for inside the cavity are obtained using the \ac{pxLDA} functional. 
The dashed lines represent the photon frequency of the photon mode, while the vertical solid black lines in both inside-cavity cases are the eigenvalues from the \ac{JC} model (please see the main text).  
}
\label{fig03}
\end{figure}

Figure~\ref{fig03}(a) compares the electron-density difference between inside and outside the cavity, with a frequency that is in resonance with the $1$st excited state of the bare hydrogen atom and the polarization along the $x$ direction. 
We evaluate this using \ac{OEP}, \ac{pxLDA}, and $\eta_{\rm c}$-\ac{px} functionals. 
Similar to the \ac{HO} case, both \ac{px} and \ac{pxLDA} approximations overestimate electron-density changes in the weak-coupling regime ($\lambda=5\times10^{-5}$), with \ac{px} results (not shown) an order of magnitude higher.
Applying a renormalization factor $\eta_{\rm{c}}=0.1$ to \ac{px} reproduces the \ac{OEP} results.
In the strong-coupling regime ($\lambda=0.5$), we provide the \ac{OEP}, \ac{px}, and \ac{pxLDA} results. 
Although \ac{OEP} may not provide accurate results due to its perturbative nature in this regime, we anticipate that \ac{px} and \ac{pxLDA} offer upper and lower bounds for the electron-density difference based on our experience with the other test systems.

After analyzing the ground-state electron density, we proceed to compute the linear-response optical absorption spectra for the hydrogen atom inside the cavity using the \ac{pxLDA} functional. 
While the \ac{OEP} functional in exchange-approximation is suitable for weak-coupling spectra~\cite{flick.welakuh.ea_2019}, exploring strong coupling with approximate \ac{QEDFT} functionals for realistic materials remains uncharted territory. 
Figure~\ref{fig03}(b) shows the optical absorption spectrum (or cross-section) of the hydrogen atom interacting with a photon mode in the strong-coupling regime. 
The polarization is along the $x$ direction, and the photon frequency is in resonance with the atom's $1$st excitation, i.e., $\omega_{\alpha} = 0.3745$.
Outside the cavity, the first peak corresponds to three dipole transitions: $1s\rightarrow 2p_{x}$, $1s\rightarrow 2p_{y}$, and $1s\rightarrow 2p_{z}$. 
These $2p$ orbitals have degenerate energy levels.
However, inside the cavity with a light-matter coupling of $\lambda = 0.05$, part of the first peak splits into two peaks.
%
These are the lower and upper polaritons arising from the hybridization between the $1s \rightarrow 2p_{x}$ transition and the photon mode. 
The other two transitions, $1s\rightarrow 2p_{y}$ and $1s\rightarrow 2p_{z}$, remain largely unchanged since the photon mode is polarized along the $x$ direction.
As the light-matter coupling increases from $0.05$ to $0.1$, the Rabi splitting between the lower and upper polariton doubles. 
Notably, other peak positions are also influenced, even when the photon mode is not in resonance with them. 
%

Here we compare our QEDFT results with the widely-used \ac{JC} model, which describes light-matter coupled two-level systems well in the weak coupling regime.
The \ac{JC} Hamiltonian within the \textit{rotating wave approximation}, describing a two-level system with energy difference $\omega_{0}$ coupled to a photon mode with photon frequency $\omega$ and polarization $\pol$, is given by (in Hartree atomic units):
\begin{equation*}
    \hat{H}_{\rm{JC}} = \omega \hat{a}^{\dagger}\hat{a} + \omega_{0}\frac{\hat{\sigma}_{z}}{2} + g(\hat{a}\hat{\sigma}_{+} + \hat{a}^{\dagger}\hat{\sigma}_{-}) + \frac{\omega_{0}}{2},
\end{equation*}
where $\hat{\sigma}_{z}$ is the Pauli matrix representing the two energy levels, $g=\lambda\sqrt{\omega/2}(\mathbf{d}\cdot\pol)$~\footnote{In SI units, $g = \mathbf{d}\cdot\pol\sqrt{\omega/2\hbar\varepsilon_{0}V} = \mathbf{d}\cdot\pol(\lambda/\hbar)\sqrt{\omega/2}$, where $\lambda = \sqrt{\hbar/\epsilon _{0}V}$ and V the mode volume} where $\mathbf{d}$ is the dipole matrix element between the two levels, and the last term ($\omega_{0}/2$) is used to reset the energy of the lowest state of the two levels system to zero.
The polaritonic eigenvalues for the photon vacuum state ($n=0$) are given by $E_{\pm}(n=0, \delta) = \frac{\omega + \omega_{0}}{2} \pm \frac{1}{2}\sqrt{4g^{2}+\delta}$ where $\delta = \omega_{0}-\omega$. 
For the $1s$ and $2p_{x}$ orbitals of the hydrogen atom coupled to a photon mode with $\delta=0$, the polariton energies are $E_{\pm}(n=0,\delta=0) = \omega_{0}\pm g$.
%
The dipole matrix element between the $1s$ and $2p_{x}$ orbital is $\langle 2p|x|1s\rangle = (\frac{2}{3})^{5}4\sqrt{2}$, given the analytical functions of the two orbitals, leading to $g = (\frac{2}{3})^{5}4\sqrt{2} \lambda \sqrt{\omega_{0}}$ in the resonance condition. 
For $\lambda=0.05$, $g = 0.01613$ ($= 0.4389$ eV). 
The polaritonic energies obtained from the JC model for this scenario are also shown in Fig.~\ref{fig03}(b).
Our \ac{QEDFT} results at a coupling parameter of $0.05$ align with those derived from the \ac{JC} model. 
However, discrepancies emerge in the strong coupling regime ($\lambda = 0.1$), where our \ac{QEDFT} results show an asymmetric Rabi splitting, a feature absent in the \ac{JC} model. 
Asymmetric Rabi splitting in the strong-coupling regime naturally arises also in other \textit{ab initio} spectra~\cite{schnappinger.sidler.ea_2023, schnappinger.kowalewski_2023}, and can also be observed and studied experimentally~\cite{canales.karmstrand.ea_2023}.

\section{Conclusion and outlook}
In summary, we have explored the reliability of \ac{px} and \ac{pxLDA} approximations, which are based on expressing the quantum fluctuations of the photons by the quantum-matter fluctuations, across various dimensional systems, coupling parameters, and frequency regimes. 
While designed for ultra-strong coupling, these non-perturbative approximations can achieve accuracy in weak-coupling situations by incorporating renormalization factors that take electron-photon correlation contributions into account. 
We demonstrated the efficiency of these methods in recovering both qualitative and quantitative exact results. 
Moreover, we established connections between changes in optical spectra and modifications in the ground state of strongly coupled light-matter systems. 
Accessing both static and time-dependent observables, such as changes in ground-state densities and optical spectra, within a unified theoretical framework opens up a lot of intriguing possibilities to explore in more detail the modifications of chemical and material equilibrium properties as uncovered by seminal experimental investigations in polaritonic chemistry and materials science~\cite{ebbesen_2016a,sanvitto.kena-cohen_2016,friskkockum.miranowicz.ea_2019,garcia-vidal.ciuti.ea_2021,schlawin.kennes.ea_2022,ebbesen.rubio.ea_2023}.
This underscores the potential of \textit{ab initio} QED methods, particularly \ac{QEDFT}, for future applications to the control of chemical and materials processes in cavities of tailored quantum environments. 
Our study demonstrates that even simple approximate functionals yield qualitatively correct results and can be systematically improved, not only qualitatively but also quantitatively, through comparative analysis.

%
%
Here we concentrated on simple finite systems to thoroughly examine the reliability of various approximate functionals. 
Unlike the often-employed unitarily-equivalent length-form of the \ac{PF} Hamiltonian~\cite{craig1998molecular,schafer.ruggenthaler.ea_2020a,mandal.taylor.ea_2023,ruggenthaler.sidler.ea_2023}, the presented approximations, grounded in the velocity gauge, offer direct applicability to extended systems, setting the stage for their application to solids. 
The velocity gauge provides a direct connection to full minimal-coupling considerations, with a complete minimal-coupling form of the \ac{px} potential already established in the literature~\cite{schafer.buchholz.ea_2021}. 
Exploring this approach for the full minimal-coupling \ac{PF} Hamiltonian within the Maxwell-Pauli-Kohn-Sham framework~\cite{jestaedt.ruggenthaler.ea_2019} and considering functionals for chiral cavities~\cite{hubener.degiovannini.ea_2021,riso.grazioli.ea_2023} represent promising future directions.

Beyond ground-state scenarios, the development of non-adiabatic (current-)density functionals is crucial. 
Leveraging the local-force equation, similar to (electron-only) time-dependent \ac{DFT}, offers a viable strategy~\cite{tchenkoue.penz.ea_2019,ruggenthaler.tancogne-dejean.ea_2022}. 
Future research will also involve extensive benchmarking of existing approximate \ac{QEDFT} functionals. 
While benchmarks for standard electronic-structure methods are well-established, reliable reference results for polaritonic systems, especially in realistic contexts, are still in progress. 
This challenge extends beyond single molecules and atoms to collectively-coupled systems~\cite{feist.galego.ea_2018,li.cui.ea_2022,sidler.ruggenthaler.ea_2022,campos-gonzalez-angulo.poh.ea_2023,mandal.taylor.ea_2023, ruggenthaler.sidler.ea_2023}, where the non-perturbative interplay between electronic and ro-vibrational degrees of freedom in large ensembles presents intriguing possibilities~\cite{schlawin.kennes.ea_2022,simpkins.dunkelberger.ea_2023,hirai.hutchison.ea_2023}. 
Exploring these interactions could unveil novel avenues for electronic-structure methods by bridging disparate energy and length scales.

\begin{acknowledgments}
\vspace{10pt}
This work was supported by the Cluster of Excellence ‘CUI:Advanced Imaging of Matter’ of the Deutsche Forschungs-gemeinschaft (DFG), EXC 2056, Grupos Consolidados (IT1453-22) and SFB925.
I-T. Lu thank Alexander von Humboldt Foundation for the support from Humboldt Research Fellowship. 
The authors thank Dr. Christian Schäfer, Dr. Johannes Flick, Dr. Davis M. Welakuh, Dr. Heiko Appel, and Iman Ahmadabadi for the fruitful discussion.
I-T. Lu thank Dr. Davis M. Welakuh for providing an in-house (private) python version of LibQED, which has been developed by previous and current members in Prof. Angel Rubio's research group, to solve the non-relativistic \ac{PF} Hamiltonian via the exact diagonalization.  
The Flatiron Institute is a division of the Simons Foundation.
\end{acknowledgments}


\appendix

\section{Derivation for the Pauli-Fierz Hamiltonian with the dressed photon modes}\label{sec:appendix-dressed-photons}

After the expansion of the kinetic energy term in Eq.~\eqref{eq:H_PF_start}, the PF Hamiltonian can be written as the sum of 1) the Hamiltonian for the matter $\hat{H}_{\rm{M}}$, 2) the Hamiltonian for the photon system $\hat{H}_{\gamma}$, and 3) the interaction between the two systems $\frac{1}{c}\hat{\bA}\cdot\hbJp$ with the paramagnetic current operator $\hat{\mathbf{J}}_{\rm{p}}=\sum_{l=1}^{N_{e}}(-i\nabla_{l})$: 
\begin{equation*}
\hat{H}_{\rm{PF}} = \hat{H}_{\rm{M}} +\frac{1}{c}\hat{\bA} \cdot \hat{\mathbf{J}}_{\rm{p}} + \hat{H}_{\gamma},
\end{equation*}
where the Hamiltonian for the matter system is defined as 
\begin{equation*}
   \hat{H}_{\rm{M}}=-\frac{1}{2}\sum_{l=1}^{N_{e}}\nabla_{l}^{2}+\sum_{l=1}^{N_{e}}v_{\rm{ext}}(\mathbf{r}_{l})+\frac{1}{2}\sum_{l\neq k}^{N_{e}}w(\mathbf{r}_{l},\mathbf{r}_{k}),
\end{equation*}
while that for the photon system is 
\begin{equation*}\label{eq:pure-photon-part}
\hat{H}_{\gamma} = \sum_{\alpha=1}^{M_{p}}\omega_{\alpha}\left(\hat{a}^{\dagger}_{\alpha}\hat{a}_{\alpha}+\frac{1}{2}\right)+\frac{N_{e}}{2c^{2}}\hat{\bA}^{2},
\end{equation*}
where only in this appendix we overload the notation $\hat{H}_{\gamma}$ for many modes, compared to the one in the main text.
%
Here we introduce a pair of harmonic coordinates for the bare photons: 
\begin{equation*}
\begin{aligned}
& \hat{q}_{\alpha}=\frac{1}{\sqrt{2\omega_{\alpha}}}\left(\hat{a}^{\dagger}_{\alpha}+\hat{a}_{\alpha}\right), \\
& \hat{p}_{\alpha}=i\sqrt{\frac{\omega_{\alpha}}{2}}\left(\hat{a}^{\dagger}_{\alpha}-\hat{a}_{\alpha}\right),
\end{aligned}
\end{equation*}
and rewrite $\hat{H}_{\gamma}$ in terms of the harmonic coordinates as 
\begin{equation}\label{eq:H_pt_harmonic}
\begin{aligned}
   \hat{H}_{\gamma}&=\frac{1}{2}\sum_{\alpha=1}^{M_{p}}\left(\hat{p}_{\alpha}^{2}+\hat{q}_{\alpha}\sum_{\alpha'=1}^{M_{p}}W_{\alpha\alpha'}\hat{q}_{\alpha'}\right)
   \\
   &=\frac{1}{2}\left(\hat{\mathbf{P}}^{\intercal}\hat{\mathbf{P}}+\hat{\mathbf{Q}}^{\intercal}\mathbf{W}\hat{\mathbf{Q}}\right),
\end{aligned}
\end{equation}
where we introduce a few notations to simplify the Hamiltonian: $\hat{\mathbf{P}} = (\hat{p}_{1},...,\hat{p}_{M_{p}})^{\intercal}$ ($\intercal$ means transpose),   $\hat{\mathbf{Q}} = (\hat{q}_{1},...,\hat{q}_{M_{p}})^{\intercal}$ , and $W_{\alpha\alpha'}=\omega_{\alpha}^{2}\delta_{\alpha\alpha'}+N_{e}\lambda_{\alpha}\lambda_{\alpha'}\pol_{\alpha}\cdot\pol_{\alpha'}$. 
The matrix $\mathbf{W}$ is real and symmetric, and can be diagonalized using an orthonormal matrix $\mathbf{U}$, such that $\mathbf{\tilde{\Omega}}=\mathbf{U}\mathbf{W}\mathbf{U^{T}}$ with eigenvalues $\tilde{\omega}_{\alpha}^{2}$, where $\tilde{\omega}_{\alpha}$ is the dressed frequency for the $\alpha$-th photon mode.
Next, we use a pair of transformed harmonic coordinates $\hat{\tilde{\mathbf{P}}} = \mathbf{U}\hat{\mathbf{P}}$ and $\hat{\tilde{\mathbf{Q}}}= \mathbf{U}\hat{\mathbf{Q}}$, that is,  $\hat{\tilde{p}}_{\alpha}=\sum_{\beta=1}^{M_{p}}U_{\alpha\beta}\hat{p}_{\beta}$ and $\hat{\tilde{q}}_{\alpha}=\sum_{\beta=1}^{M_{p}}U_{\alpha\beta}\hat{q}_{\beta}$,  respectively. 
The Hamiltonian for the photon system $\hat{H}_{\gamma}$ [Eq.~(\ref{eq:H_pt_harmonic})] becomes, with the help of the identity $\mathbf{I}= \mathbf{U}^{\intercal}\mathbf{U}$,
\begin{equation*}
\hat{H}_{\gamma}
=\frac{1}{2}\sum_{\alpha=1}^{M_{p}}\left(\hat{\tilde{p}}_{\alpha}^{2}+\tilde{\omega}^{2}_{\alpha}\hat{\tilde{q}}_{\alpha}^{2}\right)
=\sum_{\alpha=1}^{M_{p}}\tilde{\omega}_{\alpha}\left(\hat{\tilde{a}}^{\dagger}_{\alpha}\hat{\tilde{a}}_{\alpha}+\frac{1}{2}\right),
\end{equation*}
where in the second equality we define the annihilation $\hat{\tilde{a}}_{\alpha}$ and creation operator $\hat{\tilde{a}}^{\dagger}_{\alpha}$ for the dressed photons as %
\begin{equation*}
\begin{aligned}
& \hat{\tilde{a}}_{\alpha}=\frac{1}{\sqrt{2\tilde{\omega}_{\alpha}}}\left(\tilde{\omega}_{\alpha}\hat{\tilde{q}}_{\alpha}+i\hat{\tilde{p}}_{\alpha}\right), \\
& \hat{\tilde{a}}_{\alpha}^{\dagger}=\frac{1}{\sqrt{2\tilde{\omega}_{\alpha}}}\left(\tilde{\omega}_{\alpha}\hat{\tilde{q}}_{\alpha}-i\hat{\tilde{p}}_{\alpha}\right).
\end{aligned}
\end{equation*}

Since we now use the dressed photons instead of the bare photons, the vector potential operator $\hat{\bA}$ needs to be rewritten in terms of the dressed photons as well: 
\begin{equation*}
\begin{aligned}
\hat{\bA} & =c\sum_{\alpha=1}^{M_{p}}\lambda_{\alpha}\pol_{\alpha}\hat{q}_{\alpha} = c\sum_{\alpha=1}^{M_{p}}\lambda_{\alpha}\pol_{\alpha}\sum_{\beta=1}^{M_{p}}U_{\beta\alpha}\hat{\tilde{q}}_{\beta} \\ 
& = c\sum_{\alpha=1}^{M_{p}}\left[\sum_{\beta=1}^{M_{p}}U_{\alpha\beta}\lambda_{\beta}\pol_{\beta}\right]\hat{\tilde{q}}_{\alpha} 
= c\sum_{\alpha=1}^{M_{p}}\tilde{\lambda}_{\alpha}\tilde{\pol}_{\alpha}\hat{\tilde{q}}_{\alpha} = \hat{\tilde{\bA}}, 
\end{aligned}
\end{equation*}
where we define the coupling $\tilde{\lambda}_{\alpha}$ and polarization $\tilde{\pol}_{\alpha}$ for each dressed photon mode using the relation:
\begin{equation}\label{eq:new_coupling_pol}
\tilde{\lambda}_{\alpha}\tilde{\pol}_{\alpha}=\sum_{\beta=1}^{M_{p}}U_{\alpha\beta}\lambda_{\beta}\pol_{\beta}.    
\end{equation}
The polarizations for each bare and each dressed photon mode, $\pol_{\alpha}$ and $\tilde{\pol}_{\alpha}$, are normalized, i.e., $|\pol_{\alpha}|=|\tilde{\pol}_{\alpha}|=1$. 
Using this property, one can obtain the coupling $\tilde{\lambda}_{\alpha}$ for the $\alpha$-th dressed photon mode. 
If we assume that all the photon modes have the same coupling parameter, then the polarization for each dressed photon mode becomes $\tilde{\pol}_{\alpha}=\sum_{\beta=1}^{M_{p}}U_{\alpha\beta}\pol_{\beta}$, which reproduces the results in Ref.~\cite{faisal1987theory}. 
However, if the coupling parameters are different, then one should use Eq.~(\ref{eq:new_coupling_pol}) to get the correct polarization instead. 
Note that the vector potential operator does not transform via the unitary matrix but expresses itself in terms of the dressed photon modes. 
Therefore, the PF Hamiltonian [Eq.~(\ref{eq:H_PF_start})] in terms of the dressed photon modes becomes Eq.~(\ref{eq:PFH_dressed_photons}).

\section{Electron-photon exchange potential for one electron coupled to one photon mode}\label{sec:appendix-1pxc}

Here we focus on the \ac{px} potential for one effective photon mode.
For many cavity-modes cases, we can add similar \ac{px} potentials together with the corresponding light-matter coupling $\tilde{\lambda}_{\alpha}$ and dressed photon frequency $\tilde{\omega}_{\alpha}$, i.e., $v_{\rm{px}}(\br) = \sum_{\alpha=1}^{M_{p}} v_{\rm{px},\alpha}(\br)$, where $v_{\rm{px},\alpha}(\br)$ is the \ac{px} potential for the $\alpha$th photon mode. 

Assume that the wave function of the ground state $\psi_{0}(\br)=\rho^{1/2}(\br)$ is real, the px potential for one-electron cases can be obtained analytically from the Poisson equation Eq.~\eqref{eq:vpx-poisson}, together with the definition of the paramagnetic current operator $\hbJp$, the paramagnetic current density $\hbjp(\br)$, and $N_{e}=1$:
\begin{equation*}
\begin{aligned}
\nabla^{2} v_{\rm{px}}(\mathbf{r})& =-\frac{{\tilde{\lambda}}^{2}}{\tilde{\omega}^{2}}\nabla\cdot\left[\frac{(\tilde{\pol}\cdot\nabla)\langle(\tilde{\pol}\cdot\hbJp)\hbjp(\br)\rangle_{\psi_{0}}}{\rho(\br)}\right] \\
& = \frac{{\tilde{\lambda}}^{2}}{2\tilde{\omega}^{2}}\nabla^{2}\left[\frac{(\tilde{\pol}\cdot\nabla)^{2}\psi_{0}(\br)}{\psi_{0}(\br)}\right].
\end{aligned}
\end{equation*}
%
%
Thus, the px potential is 
\begin{equation*}
v_{\rm{px}}(\mathbf{r}) = \frac{{\tilde{\lambda}}^{2}}{2\tilde{\omega}^{2}} \frac{(\tilde{\pol}\cdot\nabla)^{2}\rho^{\frac{1}{2}}(\br)}{\rho^{\frac{1}{2}}(\br)}.
\end{equation*}

\section{Electron-photon exchange energy within the local density approximation}\label{sec:epxlda}

Here we derive the \ac{px} energy within the local density approximation [Eq.~\eqref{eq:epxlda-formula}] from two approaches, the reduced density matrix (RDM) and the virial relation. 

\subsection{Derivation from the reduced density matrix}

%
We follow a similar strategy to derive the \ac{px} energy within the local-density approximation as the pxLDA force derived in Ref.~\cite{schafer.buchholz.ea_2021}.
First, we define the 1-particle and 2-particle RDM for the Slater-determinant $\Phi(\br_{1},\br_{2},\br_{3},...,\br_{N_{e}})=\Phi(\br_{1},\underline{\br})=\Phi(\br_{1},\br_{2},\dunderline{\br})$, where we use $\underline{\br} = (\br_{2},\br_{3},...,\br_{N_{e}})$ and $\dunderline{\br} = (\br_{3},\br_{4},...,\br_{N_{e}})$. 
The 1-particle RDM (1RDM) is defined as
\begin{equation*}\label{eq:1RDM}
\rho_{(1)}(\br_{1},\br'_{1}) = N_{e}\int d\underline{\br}\ \Phi(\br_{1},\underline{\br})\Phi^{*}(\br'_{1},\underline{\br}),
\end{equation*}
while the 2-particle RDM (2RDM) is defined as
\begin{equation*}\label{eq:2RDM}
\begin{aligned}
& \rho_{(2)}(\br_{1},\br_{2};\br'_{1},\br'_{2}) = \\ 
& \frac{N_{e}(N_{e}-1)}{2}\times \int d\dunderline{\br}\ \Phi(\br_{1},\br_{2},\dunderline{\br})\Phi^{*}(\br'_{1},\br'_{2},\dunderline{\br}).
\end{aligned}
\end{equation*}
Using the above formula for the 1RDM and 2RDM (with the closed-shell assumption), we can write the expectation value of the current-current correlation in Eq.~\eqref{eq:def-pxenergy-functional} as
\begin{widetext}
\begin{equation*}
\begin{aligned}
& \braket{(\tilde{\pol}_{\alpha}\cdot\hbJp)\Phi}{(\tilde{\pol}_{\alpha}\cdot\hbJp)\Phi} =  N_{e}\int d\br_{1} \int d\underline{\br}\ \left[(\tilde{\pol}_{\alpha}\cdot\nabla_{1})\Phi(\br_{1},\underline{\br})\right]^{*} \left[(\tilde{\pol}_{\alpha}\cdot\nabla_{1})\Phi(\br_{1},\underline{\br})\right]\\
& + N_{e}(N_{e}-1) \int d\br_{1} \int d\br_{2} \int d\dunderline{\br} \ \left[(\tilde{\pol}_{\alpha}\cdot\nabla_{2})\Phi(\br_{1},\br_{2},\dunderline{\br})\right]^{*} \left[(\tilde{\pol}_{\alpha}\cdot\nabla_{1})\Phi(\br_{1},\br_{2},\dunderline{\br})\right]\\
& = \int d\br_{1} (\tilde{\pol}_{\alpha}\cdot\nabla_{1'}) (\tilde{\pol}_{\alpha}\cdot\nabla_{1}) \rho_{(1)}(\br_{1},\br'_{1})\Big|_{\br'_{1}=\br_{1}} + 2 \int d\br_{1} \int d\br_{2} (\tilde{\pol}_{\alpha}\cdot\nabla_{2'}) (\tilde{\pol}_{\alpha}\cdot\nabla_{1}) \rho_{(2)}(\br_{1},\br_{2};\br'_{1},\br'_{2})\Big|_{\br'_{1}=\br_{1},\br'_{2}=\br_{2}}.
\end{aligned}
\end{equation*}
\end{widetext}
%

For closed-shell Slater-determinant states, which we assume here and below, the 2RDM can be written in terms of the 1RDM as 
\begin{equation*}
\begin{aligned}
\rho_{(2)}(\br_{1},\br_{2};\br'_{1},\br'_{2}) = \frac{1}{2} & \Big[\rho_{(1)}(\br_{1},\br'_{1})\rho_{(1)}(\br_{2},\br'_{2}) \\ & -\frac{1}{2}\rho_{(1)}(\br_{1},\br'_{2})\rho_{(1)}(\br_{2},\br'_{1})\Big].
\end{aligned}
\end{equation*}
The 1RDM for the 
\ac{HEG}, which we assume here, is given as 
%
\begin{equation*}
 \rho_{(1)}(\br_{1},\br'_{1}) = \frac{2}{(2\pi)^{d}}\int_{|\bk|< k_{\rm{F}}} d\bk\ e^{i\bk\cdot(\br_{1}-\br'_{1})},
\end{equation*}
where $k_{\rm{F}}(\br) = 2\pi[\rho(\br)/2V_{d}]^{1/d}$.
Within the \ac{HEG} approximation, we have 
\begin{equation*}
    \nabla_{1}\rho_{(1)}(\br_{1},\br'_{1})\Big|_{\br_{1},\br'_{1}} = \frac{2i}{(2\pi)^{d}}\int_{|\bk|< k_{\rm{F}}}d\bk\ \bk = 0.
\end{equation*}
Therefore, the expectation value within LDA becomes
%
\begin{equation*}
\begin{aligned}
\langle{(\tilde{\pol}_{\alpha}\cdot\hbJp)^{2}}\rangle
& = \int d\br_{1} \frac{2}{(2\pi)^{d}}\int_{|\bk|< k_{\rm{F}}}d\bk\ (\tilde{\pol}_{\alpha}\cdot\bk)^{2} \\ & -\int d\br_{1} \frac{2}{(2\pi)^{d}}\int_{|\bk|< {\rm{min}}(k_{\rm{F}},k'_{\rm{F}})}d\bk\ (\tilde{\pol}_{\alpha}\cdot\bk)^{2}.
\end{aligned}
\end{equation*}
%
For the HEG, $k'_{\rm{F}}=k_{\rm{F}}$, the expectation value vanishes as expected. 
For an inhomogeneous medium, min($k_{\rm{F}},k'_{\rm{F}}$) approaches zero because $k'_{\rm{F}}$ can possibly get small.
Therefore, we propose the following formula for the px energy within the LDA approximation:
\begin{equation*}
    E_{\rm{pxLDA}} = \int d\br \left(\sum_{\alpha=1}^{M_{p}}\frac{-\tilde{\omega}_{\alpha}^{2}}{2\tilde{\omega}_{\alpha}^{2}}\right)\left[\frac{2\kappa}{(2\pi)^{d}}\int_{|\bk|< k_{\rm{F}}}d\bk\ (\tilde{\pol}_{\alpha}\cdot\bk)^{2}\right],
\end{equation*}
where we introduce a factor $\kappa\in [0,1]$ where $\kappa=0$ for the HEG and $\kappa=1$ for the maximally inhomogeneous limit to take all situations into account.
We use the maximally inhomogeneous limit ($\kappa = 1$) in this work. 
One can recover other scenarios by including the factor $\kappa$.
The square bracket in the formula for the pxLDA energy can be evaluated in the polar coordinates in $d$-dimension:
\begin{equation*}
\frac{2}{(2\pi)^{d}}\int_{|\bk|< k_{\rm{F}}}d\bk\ (\tilde{\pol}_{\alpha}\cdot\bk)^{2} = \frac{2 V_{d}}{(2\pi)^{d}}\frac{k_{\rm{F}}^{d+2}(\br)}{d+2}.
\end{equation*}
The px energy within the LDA becomes
\begin{equation}\label{eq:epx-LDA}
E_{\rm{pxLDA}} =  \left(\sum_{\alpha=1}^{M_{p}}\frac{-\tilde{\lambda}_{\alpha}^{2}}{\tilde{\omega}_{\alpha}^{2}}\right)\frac{2\pi^{2}}{d+2}\left(\frac{1}{2V_{d}}\right)^{\frac{2}{d}}\int d\br\  \rho^{\frac{2+d}{d}}(\br).
\end{equation}
Using the above pxLDA energy, we can obtain the \textit{isotropic} pxLDA potential via the functional derivative with respect to the density as
\begin{equation*}
v_{\rm{pxLDA}}^{\rm{iso}}(\br) = \left(\sum_{\alpha=1}^{M_{p}}\frac{-2\pi^{2}\tilde{\lambda}_{\alpha}^{2}}{d\tilde{\omega}_{\alpha}^{2}}\right)\left[\frac{\rho(\br)}{2V_{d}}\right]^{\frac{2}{d}}.
\end{equation*}
Note that if we take the above isotropic pxLDA potential into the KS equations, we would lose the information of the polarization direction of those photon modes when solving the KS equations. 
However, if we take the pxLDA potential obtained from the force balance equation to compute the density, then we compute the pxLDA energy using Eq.~\eqref{eq:epx-LDA}, which implicitly contains the information of the polarizations of those photon modes. 
%
%

\subsection{Derivation from the virial relation}

The electron-photon exchange force (and its LDA version) is 
\begin{equation*}
    \mathbf{F}_{\rm{px(LDA)}}(\br) = \sum_{\alpha=1}^{M_{p}}\frac{\tilde{\lambda}_{\alpha}^{2}}{\tilde{\omega}_{\alpha}^{2}}(\tilde{\pol}_{\alpha}\cdot\nabla)\mathbf{f}_{\alpha,\rm{px(LDA)}}(\br),
\end{equation*}
where $\mathbf{f}_{\alpha,\rm{pxLDA}}(\br)$ [Eq.~\eqref{eq:sub-pxforce}] is rewritten in terms of electron density:
\begin{equation*}
    \mathbf{f}_{\alpha,\rm{pxLDA}}(\br) = \frac{(2\pi)^{2}}{d+2}\rho(\br)\left[\frac{\rho(\br)}{2V_{d}}\right]^{\frac{2}{d}}\tilde{\pol}_{\alpha}.
\end{equation*}
The pxLDA energy can be obtained using the virial relation (with a factor of $1/2$ due to the photon-energy counter term) as 
\begin{equation*}
\begin{aligned}
    & E_{\rm{pxLDA}} = \frac{1}{2}\int d\br \ \br\cdot\mathbf{F}_{\rm{pxLDA}}(\br) \\
    & = \frac{1}{2}\frac{(2\pi)^{2}}{d+2}\sum_{\alpha=1}^{M_{p}}\frac{\tilde{\lambda}_{\alpha}^{2}}{\tilde{\omega}_{\alpha}^{2}}\int d\br \  (\tilde{\pol}_{\alpha}\cdot\br)(\tilde{\pol}_{\alpha}\cdot\nabla) \left[\rho(\br)\left(\frac{\rho(\br)}{2V_{d}}\right)^{\frac{2}{d}}\right] \\ 
    & =  \left(\sum_{\alpha=1}^{M_{p}}\frac{-\tilde{\lambda}_{\alpha}^{2}}{\tilde{\omega}_{\alpha}^{2}}\right)\frac{2\pi^{2}}{d+2}\left(\frac{1}{2V_{d}}\right)^{\frac{2}{d}}\int d\br\  \rho^{\frac{2+d}{d}}(\br),
\end{aligned}
\end{equation*}
where we use integration by parts in the last line. 
The px energy obtained from the virial relation is the same as the one obtained from the RDM approach.

Here is a side note regarding the total \ac{px} force (and its LDA version): it satisfies the zero-force condition, i.e.,
\begin{equation}\label{eq:zero-pxlda-force}
\begin{aligned}
    & \int_{\Omega} d\br\ \mathbf{F}_{\rm{px}}(\br) = \sum_{\alpha=1}^{M_{p}} \frac{\tilde{\lambda}_{\alpha}^{2}}{\tilde{\omega}_{\alpha}^{2}} \int_{\Omega} d\br \ (\tilde{\pol}_{\alpha}\cdot\nabla) \mathbf{f}_{\alpha,\rm{px}}(\br) \\
    & = \sum_{\alpha=1}^{M_{p}} \frac{\tilde{\lambda}_{\alpha}^{2}}{\tilde{\omega}_{\alpha}^{2}} \int_{\Omega} d\br\ \left[\nabla (\tilde{\pol}_{\alpha}\cdot\mathbf{f}_{\alpha,\rm{px}}(\br))-\tilde{\pol}_{\alpha}\cross(\nabla\cross\mathbf{f}_{\alpha,\rm{px}}(\br))\right] \\
    & = \sum_{\alpha=1}^{M_{p}} \frac{\tilde{\lambda}_{\alpha}^{2}}{\tilde{\omega}_{\alpha}^{2}} \left[\int_{\partial\Omega} d\mathbf{S} \ \tilde{\pol}_{\alpha}\cdot\mathbf{f}_{\alpha,\rm{px}}(\br) - \tilde{\pol}_{\alpha}\cross \int_{\partial\Omega} d\mathbf{S}\cross\mathbf{f}_{\alpha,\rm{px}}(\br) \right]\\
    & = 0,
\end{aligned}
\end{equation}
where $\Omega$ is the volume of interest and its surface $\partial \Omega$.
In the second line of Eq.~\eqref{eq:zero-pxlda-force}, we use the vector calculus identity $\nabla(\mathbf{A}\cdot\mathbf{B}) = (\mathbf{A}\cdot\nabla)\mathbf{B} + (\mathbf{B}\cdot\nabla)\mathbf{A} + {\mathbf{A}}\cross(\nabla\cross{\mathbf{B}}) + \mathbf{B}\cross(\nabla\cross\mathbf{A})$ where $\mathbf{A} = \tilde{\pol}_{\alpha}$ and $\mathbf{B} = \mathbf{f}_{\alpha,\rm{px}}(\br)$.
In the third line of Eq.~\eqref{eq:zero-pxlda-force}, we use the following two identities: 1) $\int_{\Omega}d\br\ \nabla\phi(\br) = \int_{\partial \Omega}d\mathbf{S}\ \phi(\br)$, where $\phi(\br)$ is a scalar function, and 2) $\int_{\Omega}d\br\ \nabla\cross\mathbf{A}(\br) = \int_{\partial \Omega}d\mathbf{S}\cross\mathbf{A}(\br)$, where $\mathbf{A}$ is a vector-valued function.
In the last line Eq.~\eqref{eq:zero-pxlda-force}, the surface integrals vanish in both finite and periodic systems.
More detailed discussions on subtle virial-relation issues can be found in works addressing force-based functionals in standard electron-only density-functional theory, such as Ref~\cite{ruggenthaler.tancogne-dejean.ea_2022}.

\section{Computational details}\label{sec:comp-details}

For the exact diagonalization method, the ground-state electron density $\rho^{\rm{exact}}(\br)$ is obtained by tracing out the photon-Fock space, $\rho^{\rm{exact}}(\br)=\sum_{n=1}^{n_{\rm{max}}}\Psi_{0}^{*}(\br,n)\Psi_{0}(\br,n)$, where $\Psi_{0}(\br,n)$ is the ground-state wave function of the \ac{PF} Hamiltonian with one photon mode [Eq.~\eqref{eq:PF-one-mode}], $n$ the number of photon-Fock states, and $n_{\rm{max}}$ the maximum number of Fock states we choose to converge the energies of the ground state and a few excited states for tunable light-matter couplings $\lambda$.
%
For the \ac{HO}, we use a $301$-point grid centered at the origin with a grid spacing of $\Delta x=0.07$ Bohr. 
In the case of the quantum ring, we use a $61\times61$ grid centered at the origin with a step size of $\Delta x=0.2$ Bohr. 
To achieve convergence, we use $n_{\rm{max}} = 20$ and apply a fourth-order finite difference scheme for the real-space derivatives on the grid, including the Laplacian operator for kinetic energy.

The \ac{pxLDA} functional [Eq.~\eqref{eq:pxlda-d-dimension}], together with the one-electron \ac{px} [Eq.~\eqref{eq:vpx-one-electron}] and the renormalization factors $\eta_{\rm{c}}$, are implemented in the open-source code, \textit{Octopus}~\cite{tancogne-dejean.oliveira.ea_2020}. 
%
%
%
%
In \ac{QEDFT} approaches, we use specific real-space grid sizes and box dimensions for different systems: \ac{HO} (grid size $\Delta x=0.07$ Bohr, length $21$ Bohr), quantum ring (grid size $\Delta x=0.2$ Bohr, length $20$ Bohr), and the hydrogen atom (grid size $\Delta x=0.24$ Bohr, radius $20$ Bohr).
For the \ac{OEP} functional in exchange approximation, we solve the \ac{KS} equation with the potential obtained from solving the full \ac{OEP} equation~\cite{flick.schafer.ea_2018}.

In \ac{QEDFT}, we self-consistently solve the \ac{KS} equation for the ground state. 
To calculate the optical spectrum for the hydrogen case using \textit{Octopus}, we employ time-dependent techniques, propagating the ground-state wave function with the \ac{KS} Hamiltonian, i.e., Eq.~\eqref{eq:HKS} with the adiabatic \ac{KS} potential approximations. 
The optical spectrum is obtained by Fourier-transforming the time-dependent dipole moment, which is computed using the delta-kick method with a kick strength of $0.01$/$\AA$. 
The time propagation extends for $50$ fs ($2067$ atomic units) with a time step of approximately $0.0019$ fs ($0.08$ atomic unit).

\section{A many-electron case: He atom inside a cavity}\label{sec:He-example}

To simulate a system with many electrons inside a cavity using the \ac{QEDFT} approach, we require an approximate mean-field exchange-correlation potential for the \ac{KS} potential, given by $v_{\rm{KS}}(\br) = v_{\rm{ext}}(\br) + v_{\rm{Mxc}}(\br)$. 
%
%
The approximated mean-field exchange-correlation potential $v_{\rm{Mxc}}(\br)$ consists of the Hartree potential $v_{\rm{H}}(\br)$, the electron-electron exchange-correlation potential $v_{\rm{xc}}(\br)$, and the electron-photon exchange potential $v_{\rm{px}}(\br)$, that is, $v_{\rm{Mxc}}(\br)\approx v_{\rm{H}}(\br) + v_{\rm{xc}}(\br) + v_{\rm{px}}(\br)$. 
Various methods~\cite{dreizler2012density,martin_2020}, such as \ac{LDA}, generalized gradient approximation, and \ac{OEP}, can be used to approximate and compute $v_{\rm{xc}}(\br)$, while $v_{\rm{px}}(\br)$ can be approximated using \ac{OEP}~\cite{flick.schafer.ea_2018} or the method developed in this work.
%
%
The external potential $v_{\rm{ext}}(\br)$, representing the interaction between nuclei and electrons, is typically modeled using pseudopotentials~\cite{martin_2020}. 

\begin{figure}[t]
\centering
\includegraphics[width=\linewidth]{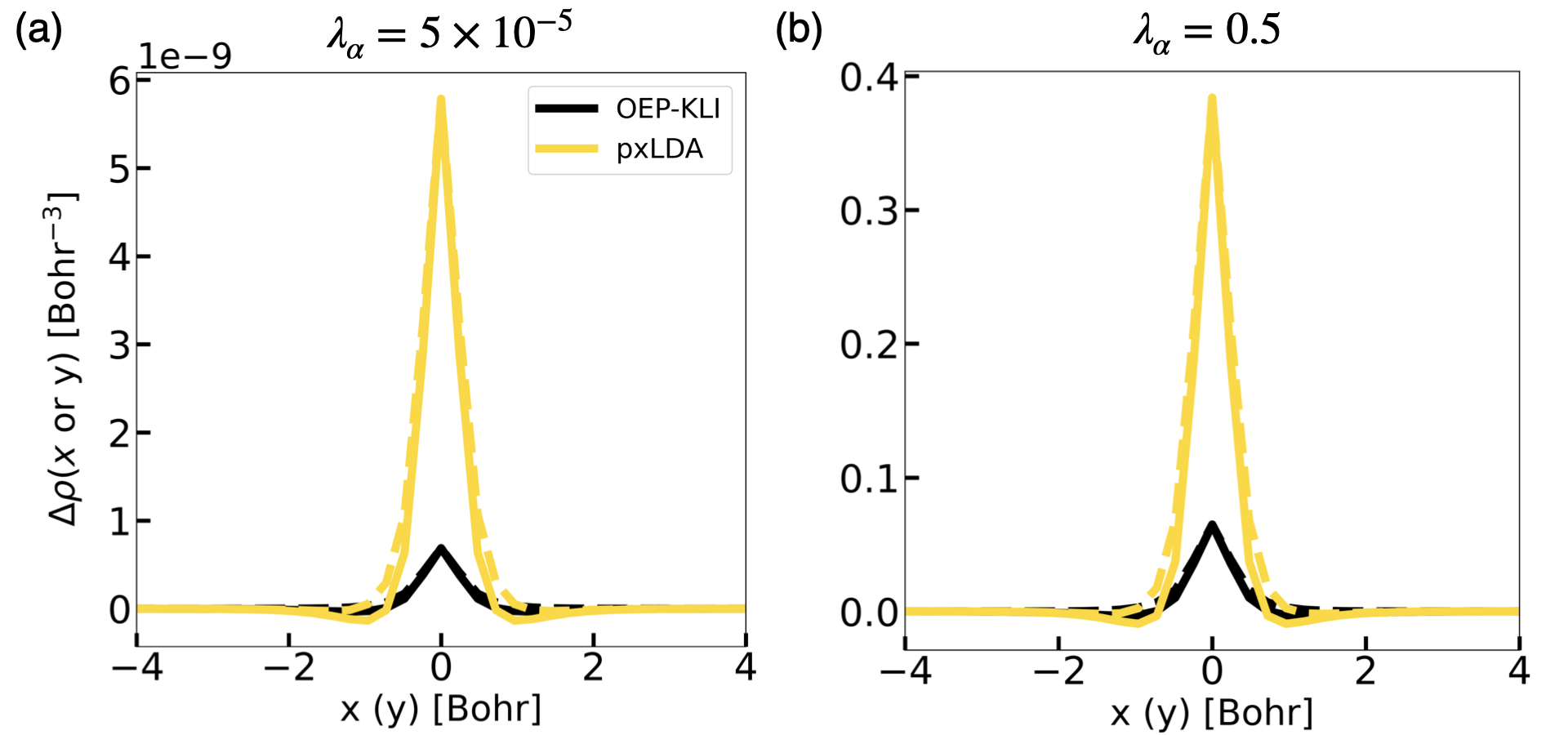}
\caption{
The photon mode is polarized along the $x$ direction and a photon frequency of $0.83927$ Hartree, in resonance with the transition between the $1s$ and $2p$ orbitals of the He atom outside the cavity.
a) and b) The electron-density differences $\Delta\rho$ along the x (solid lines) and y (dashed lines) directions in the weak- and strong-coupling regimes, computed using the OEP-KLI and pxLDA approximations for the electron-photon interaction. 
The electron-electron exchange potential is approximated and solved using the OEP-KLI approach, while the electron-electron correlation potential is approximated and obtained using the LDA approach. 
}
\label{fig04}
\end{figure}

We illustrate our approach using a helium (He) atom as a representative example of many-electron systems within an optical cavity. 
The He atom is treated with the \textit{Octopus} open-source code~\cite{tancogne-dejean.oliveira.ea_2020}, using a radius of $30$ Bohr and a real-space grid size of $0.24$ Bohr. 
The Hartwigsen-Goedecker-Hutter \ac{LDA} pseudopotential~\cite{hartwigsen.goedecker.ea_1998} models the interaction of the valence electrons with the nuclei, while the electron-electron interaction among the valence electrons is handled using the \ac{OEP} method with the \ac{KLI} approximation for the exchange potential~\cite{krieger.li.ea_1990} and \ac{LDA} for the correlation potential~\cite{perdew.zunger_1981a}.
Our primary focus is on the electron-photon interaction, specifically the electron-photon exchange potential.
We use the \ac{OEP} within the \ac{KLI} approximation as a reference~\footnote{The \ac{OEP}-\ac{KLI} results are close to the \ac{OEP} results in this case.} for realistic systems in the weak coupling regime~\cite{flick.schafer.ea_2018}, comparing it with our \ac{pxLDA} potential.
Figure~\ref{fig04} shows the electron-density difference of the He atom inside and outside the cavity for both small and large light-matter coupling.
In the weak coupling regime, our \ac{pxLDA} results show an overestimation of electron density compared to the \ac{OEP} method, akin to the hydrogen case.
%
%
This example underscores the versatility of our \ac{pxLDA} approach in extending to many-electron systems, incorporating the electron-electron exchange-correlation potential in the \ac{KS} potential.

\bibliography{references}

\end{document}